# Fundamental and Progress of $Bi_2Te_3$-based Thermoelectric Materials

Min Hong (洪敏)[1,2], Zhi-Gang Chen (陈志刚)[1,2*], and Jin Zou (邹进)[2,3*]

[1]*Centre of Future Materials, the University of Southern Queensland, Springfield, Queensland 4300, Australia*
[2]*Materials Engineering, University of Queensland, Brisbane, Queensland 4072, Australia*
[3]*Centre for Microscopy and Microanalysis, University of Queensland, Brisbane, Queensland 4072, Australia*
*Corresponding author E-mail: zhigang.chen@usq.edu.au, j.zou@uq.edu.au

Thermoelectric materials, enabling the directing conversion between heat and electricity, are one of the promising candidates for overcoming environmental pollution and the upcoming energy shortage caused by the over-consumption of fossil fuels. $Bi_2Te_3$-based alloys are the classical thermoelectric materials working near room temperature. Due to the intensive theoretical investigations and experimental demonstrations, significant progress has been achieved to enhance the thermoelectric performance of $Bi_2Te_3$-based thermoelectric materials. In this review, we first explored the fundamentals of thermoelectric effect and derived the equations for thermoelectric properties. On this basis, we studied the effect of material parameters on thermoelectric properties. Then, we analyzed the features of $Bi_2Te_3$-based thermoelectric materials, including the lattice defects, anisotropic behavior and the strong bipolar conduction at relatively high temperature. Then we accordingly summarized the strategies for enhancing the thermoelectric performance, including point defect engineering, texture alignment, and band gap enlargement. Moreover, we highlighted the progress in decreasing thermal conductivity using nanostructures fabricated by solution grown method, ball milling, and melt spinning. Lastly, we employed modeling analysis to uncover the principles of anisotropy behavior and the achieved enhancement in $Bi_2Te_3$, which will enlighten the enhancement of thermoelectric performance in broader materials.

## 1 Introduction

The rising demand for energy supply, the elimination of greenhouse gas due to carbon-based energy sources, and the enhancement in the energy consumption efficiency have sparked significant research into alternative energy sources and energy harvesting technologies. One of the promising candidates is thermoelectricity, in which heat is transferred directly into electricity.[1-3] Because of the distinct advantages of thermoelectric devices: no moving parts, long steady-state operation period, zero emission, precise temperature control and capable of function in extreme environment,[4-6] the prospect of thermoelectric applications is promising, especially for power generation and refrigeration. For the power generation mode, energy is captured from waste, environmental, or mechanical sources, and converted into an exploitable form — electricity by thermoelectric devices.[7-10] Thermoelectric materials are also able to generate power by using solar energy to create a temperature difference across thermoelectric materials.[11,12] Nuclear reactors and radioisotope thermoelectric generators can be used as spacecraft propulsion and for power supply.[13-16] For the refrigeration mode, micro thermoelectric cooling modules can be installed in the integrated circuit to tackle the heat-dissipation problem, and flexible thermoelectric materials can be equipped in the uniform of people working in the extreme environment to serve as the wearable climate control system.[17]

The wide application of thermoelectric materials requires high energy conversion efficiency and the enhancement involves the simultaneous management of several parameters. First, Seebeck coefficient ($S$), the generated voltage over the applied temperature, is defined. Since thermoelectric materials involve charge carrier transport and heat flow, electrical conductivity ($\sigma$) and thermal conductivity ($\kappa$) should be taken into account. To ensure a high output power, both $S$ and $\sigma$ should be as large as possible. To maintain the temperature difference across the material, a low $\kappa$ is preferred. Accordingly, the dimensionless figure-of-merit, $zT$ has been defined to quantify the thermoelectric performance at a given temperature ($T$), namely $zT = S^2\sigma T/\kappa$.[1,18,19] The criteria of good thermoelectric materials are high power-factor ($S^2\sigma$) and low $\kappa$. $S^2\sigma$ can be enhanced by resonant state doping,[1,20,21] minority carrier blocking,[22] band convergence,[23-27] reversible phase transition,[28-30] and quantum confinement,[31,32] and $\kappa$ can be reduced by nanostructuring,[33-39] hierarchical architecturing,[40-42] producing dislocations,[43-45] introducing stacking faults,[46,47] and incorporating nanoprecipitates into the matrix.[48-53]

Owing to the intensive theoretical studies and experimental demonstrations over the past decades, $zT$ has been improved dramatically. Fig. 1a and b plot the achieved state-of-the-art $zT$ values for both $n$-type and $p$-type materials working over a wide temperature range. $Bi_2Te_3$, PbTe, and GeSi are the classical thermoelectric materials working near room-, mid-, and high-temperature, respectively.[5,54] Through band alignment, Na-doped $PbTe_{1-x}Se_x$ exhibited $zT$ up to 1.8.[23] In addition, nanostructuring successfully improved $zT$ for GeSe to over 1.[55] Moreover, SnTe, as an environmentally friendly lead-free thermoelectric candidate, exhibited $zT$ over 1.2 through producing band convergence and introducing nanoprecipitates.[56-61] The lead-free SnSe with earth-abundant elements has attracted wide attention due to the reported record-high $zT$ of 2.6 along the $b$ axis of the single crystal.[28] Later on, Na-doping was employed to significantly enhance the mid-temperature $zT$ for SnSe single crystal.[62-66]

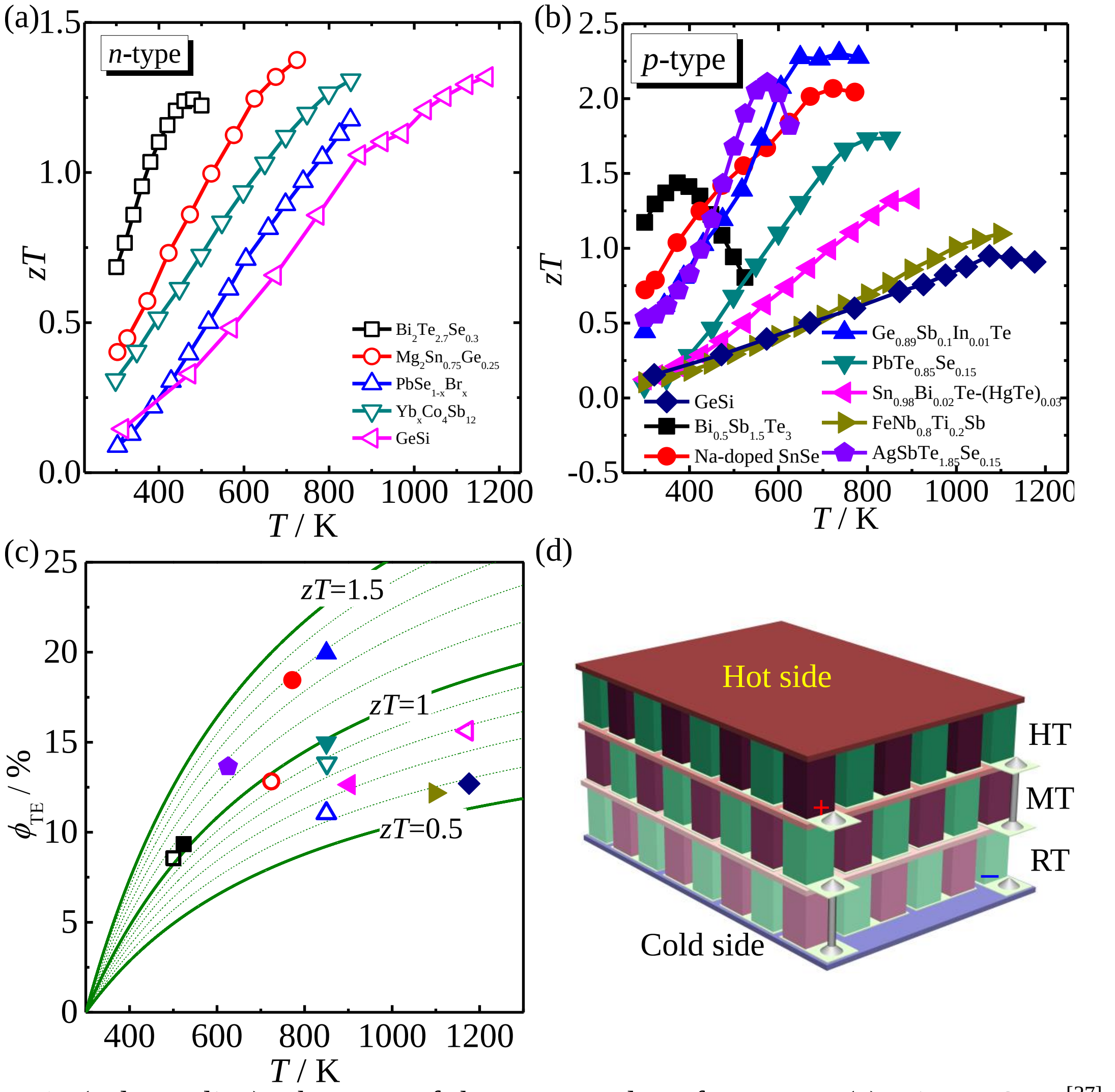

Fig. 1. (color online) The state-of-the-art *zT* values for *n*-type (a) $Bi_2Te_{2.7}Se_{0.3}$,[37] $Mg_2Sn_{0.75}Ge_{0.25}$,[25] $PbSe_{1-x}Br_x$,[67] $Yb_xCo_4Sb_{12}$,[68] and GeSi;[69] *p*-type (b) $Ge_{0.89}Sb_{0.1}In_{0.01}$,[47] $Bi_{0.5}Sb_{1.5}Te_3$,[70] $TePbTe_{0.85}Se_{0.15}$,[23] $Sn_{0.98}Bi_{0.02}Te\text{-}(HgTe)_{0.03}$,[60] $FeNb_{0.8}Ti_{0.2}Sb$,[71] $AgSbTe_{1.85}Se_{0.15}$,[46] and GeSi.[55] (c) Corresponding thermoelectric efficiency calculated using $zT_{avg}$ values of these summarized materials. (d) The schematic diagram of a multi-layer thermoelectric device assembled from materials working at high (HT), mid (MT), and room (RT) temperatures along the thermal gradient from the hot side to the cold side.

The efficiency ($\phi_{TE}$) of a thermoelectric device as a function of *zT* is calculated by[72,73]

$$\phi_{TE} = \frac{W}{Q_H} = \phi_C \left( \frac{\sqrt{1+zT} - 1}{\sqrt{1+zT} + T_C/T_H} \right) , \tag{1}$$

and $\phi_C$ is the Carnot efficiency, given by

$$\phi_C = \frac{T_H - T_C}{T_H} . \tag{2}$$

In the above equations, $Q_H$ is the net heat flow rate, $W$ is the generated electric power, $T_C$ is the cold side temperature, and $T_H$ is the hot side temperature, respectively.

According to Equation (1), the dependence of thermoelectric efficiency on average $zT$ ($zT_{avg}$) for single-leg state-of-the-art thermoelectric materials is plotted in Fig. 1c, in which the cold side temperature is set as 300 K, and the hot side temperature is the corresponding temperature for peak $zT$. The efficiency of thermoelectric materials for room-temperature power generation is lower than 10%, for mid-temperature applications, the peak efficiency could be ~ 25%, and for high-temperature applications, the efficiency is 22%.

To allow a maximum energy harvesting efficiency from a heat source, a thermoelectric device can be assembled by triple layers of *p*-*n*-junction arrays in a tandem mode, as shown in Fig. 1d. The first layer is high-temperature thermoelectric materials, while the second and third layers are respectively built from mid- and low-temperature candidates as regenerative cycles. So far, $Bi_2Te_3$ families are still the most promising room-temperature thermoelectric candidates. It is necessary to dedicate significant research into this classical thermoelectric category.

In this review, we provide an overview of the development of $Bi_2Te_3$-based thermoelectric materials, including the fundamentals on thermoelectric effects, current research progress, and new trends in $Bi_2Te_3$-based thermoelectric materials. Firstly, we study the fundamentals of thermoelectric effects and find that thermoelectric effects are substantially the re-distribution of electrons disturbed by temperature difference or extra potential and the attendant energy exchange between the free charge carriers and environment. Also, we derived the model of electronic transport coefficients from the Boltzmann equation (BTE). On this basis, we performed simulations of the effect of materials' parameters on electronic transport coefficient. Secondly, we summarized the features of $Bi_2Te_3$-based thermoelectric materials and the correspondingly developed strategies for enhancing their thermoelectric performance. Finally, we quantitatively analyzed the reported data for $Bi_2Te_3$-based thermoelectric materials, which is believed to provide innovative directions for developing high-performance thermoelectric candidates in broader materials.

## 2 Thermoelectric effect

Thermoelectric effect includes Seebeck effect and Peltier effect.[74] Seebeck effect originates from the charge current flow driven by the temperature difference. Reversibly, the charge current flow driven by extra applied voltage can generate temperature difference across the material, which is Peltier effect.[75,76] In this regards, thermoelectric effect involves how temperature difference or extra voltage disturb the distribution of free charge carriers and the attendant energy exchange.[77]

At equilibrium, the distribution of free charge carriers, for instance, electrons, follows the Fermi-Dirac function, *i.e.*

$$f_0 = \frac{1}{\exp(\frac{E-u}{k_B T})+1}\ , \tag{3}$$

in which $E$ is the energy level, $u$ is the chemical potential, $k_B$ is the Boltzmann constant, and $T$ is temperature. Generally, $u$ equals Fermi level ($E_f$).[78,79] If $E >> u$, $f_0 = 0$, and if $E << u$, $f_0 = 1$. At 0 K, the transition of $f_0$ from 0 to 1 exactly occurs at $u$, while at non-zero K the transition of $f_0$ from 0 to 1 occurs over an energy window of a few $k_BT$.[78] At a given $E$, $f_0$ is determined by both $T$ and $u$. Therefore, the temperature and external potential (*i.e.* voltage) can change $f_0$ and consequently result in the re-distribution of charge carriers, ultimately leading to the charge current flow. In this context, we present a detailed discussion on the principle of thermoelectric effects.

Fig. 2a shows a single-leg thermoelectric device consisting of a channel (thermoelectric material) with two contacts. $T_i$, $u_i$, and $f_0^i$ are temperatures, chemical potentials, and Fermi-Dirac functions for the two contacts (i = 1 and 2 representing the left and right contacts), respectively. The competition between $f_0^1$ and $f_0^2$, *i.e.* $f_0^1 - f_0^2$, accounts for the charge current flow through the channel. Without temperature or potential disturbances, (*i.e.* $T_1 = T_2$ and $u_1 = u_2$), $f_0^1 - f_0^2 = 0$. When $T_1 > T_2$ but $u_1 = u_2$ (namely, applying a temperature difference) the transition of $f_1$ from 0 to 1 occurs over a wider energy window than $f_2$ does, which means at an energy higher than the chemical potential, $f_0^1 - f_0^2 > 0$, otherwise $f_0^1 - f_0^2 < 0$. Electrons with energy higher than the chemical potential flow from the left contact to the right contact, while electrons with lower energy flow from the right contact to the left contact. Despite $f_0$, the other factor affecting the number of electrons at a certain energy is the density of states (DOS, $g(E)$), which describes the possible states of electrons at the energy level of $E$.[80] In Fig. 2a, the plotted $g(E)$ corresponds to *n*-type semiconductors and increases with increasing energy level, indicative of more electron states at the higher energy level. Consequently, the net electron current flows from hot side to cold side, resulting in the hot side to be positive and the cold side to be negative. Thus, $u_1$ becomes lower than $u_2$, and the difference between $u_1$ and $u_2$ increases when more and more electrons accumulate on the code side. Note that negative electrode has a higher potential for electrons. Fig. 2b describes the updated $f_0^1 - f_0^2$ at equilibrium, and there is no net electron flow.

For the Peltier effect, the temperature difference is generated by charge current flow, which is driven by the external voltage. Fig. 2c is the schematic illustrating the Peltier effect, in which an external voltage is applied to the two contacts, for example, left contact being negative and right contact being positive, *i.e.*, $u_1 > u_2$ for electrons, leading to the electron current flows from left to right. For simplicity, the channel is regarded as an elastic resistor, in which the electrons current does not lose energy and energy exchange only occurs at the two contacts. For the right contact, the arrival electrons with a higher energy of $E_e$ should release energy by an amount of $Q_{rel} = E_e - u_2$ per each electron. On the other hand, the provided electrons from the left contact should absorb energy to climb to the higher energy level of $E_e$ and the required energy value is $Q_{abs} = E_e - u_1$ per each electron. Accordingly, the left contact is cooled down, resulting in $T_1 < T_2$. Therefore, $f_0^1 - f_0^2$ varies, as shown in Fig. 2d. Finally, there is no net electron flow at the equilibrium. Note that within the framework of elastic resistor, $Q_{rel} - Q_{abs} = u_1 - u_2$, which is the potential difference provided by the external power; however, the practical thermoelectric materials are inelastic resistor because of the inelastic electron-electron and electron-nucleus collisions, which means electron current loses energy inside the thermoelectric materials in the form of heat, resulting in $Q_{rel} - Q_{abs} < u_1 - u_2$, namely the input power is higher than the generated heat energy difference.

Based on above discussion, charge current flow results from the disturbance of Fermi-Dirac function, which can be motivated by the temperature difference and the external potential. Seebeck effect is the Fermi-Dirac function disturbed by the temperature difference. Working under the power generation mode, the cold side corresponds to the negative pole of *n*-type, while the hot side corresponds to the negative pole of *p*-type. Peltier effect is the energy exchange between the free electrons and the contacts when Fermi-Dirac function is disturbed by an external potential. Working under the refrigeration mode, the negative pole is cooled down for *n*-type, while positive pole is cooled down for *p*-type.

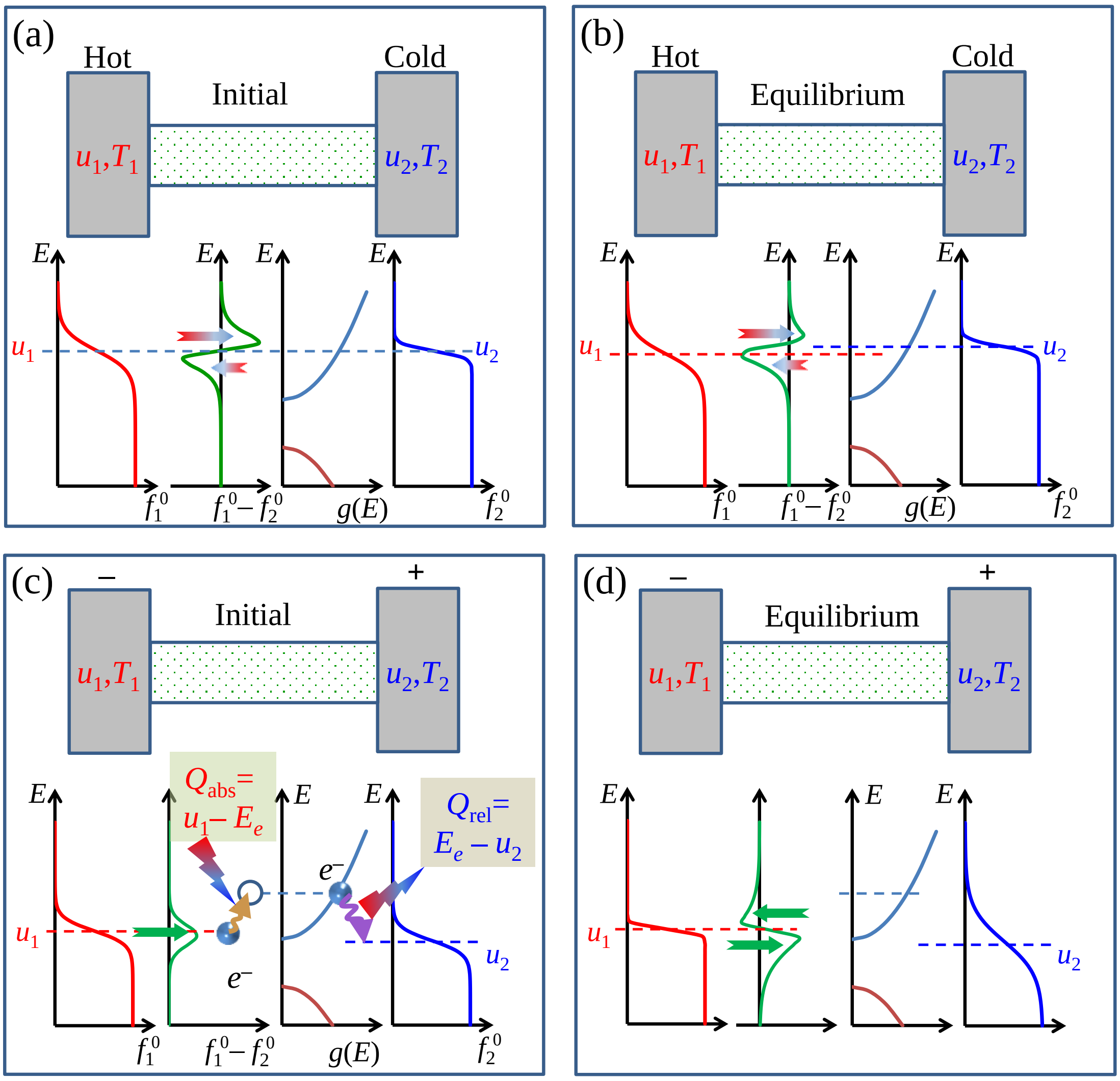

Fig. 2. (color online) Seebeck effect with charge current driven by the temperature difference for *n*-type thermoelectric materials: (a) the initial stage and (b) equilibrium. (c) Peltier effect with charge current driven by the voltage for *n*-type thermoelectric materials: (c) the initial stage and (d) equilibrium.

## 3 Equations of electronic transport coefficients from Boltzmann equation (BTE)

### 3.1 Definition of electronic transport coefficients

Thermoelectric effect involves charge current flow and heats current flow. Noteworthy, the heat current specifically refers to the thermal energy transported by charge current. By definition, charge current density ($J$), and heat current density ($Q$) are respectively expressed by[32]

$$J \equiv nev = e\int_{-\infty}^{+\infty} v(E)g(E)\left[f(E)-f_0(E)\right]dE \tag{4}$$

$$Q \equiv n(E-u)v = \int_{-\infty}^{+\infty} v(E)(E-u)g(E)\left[f(E)-f_0(E)\right]dE \tag{5}$$

in which, $n$ is the carrier concentration, $e$ is the free electron charge, and $v$ is the carrier velocity.

On this basis, $\sigma$, $S$, and the electrical thermal conductivity ($\kappa_e$) can be defined as[80]

$$\sigma \equiv \frac{J}{\varepsilon}\bigg|_{\nabla_x T=0} \tag{6}$$

$$S \equiv \frac{\varepsilon}{\nabla_x T}\bigg|_{J=0} \tag{7}$$

$$\kappa_e \equiv -\frac{Q}{\nabla_x T}\bigg|_{J=0} \tag{8}$$

with $\varepsilon$ denoting the intensity of the electrical field. Equation (6) is based on the Ohm's law without a temperature gradient. Equation (7) and (8) is the thermoelectric generator working under an open circuit. The negative sign in Equation (8) means that thermal energy is lost.

### 3.2 Steady state solution to Boltzmann Equation

From Equations (4) to (8), the derivations of $S$, $\sigma$, and $\kappa_e$ are actually to determine $f(E) - f_0(E)$ in Equations (4) and (5), which is the steady-state solution to the Boltzmann Equation (BTE).[81,82] BTE is

$$\left(\frac{\partial f}{\partial t}\right)_{field} + \left(\frac{\partial f}{\partial t}\right)_{collision} = 0 , \tag{9}$$

in which $f$, as functions of time ($t$), wave vectors (k) and position vectors (x), is the non-equilibrium distribution function of electrons. The first term of Equation (9) represents how the external fields (including electrical and temperature fields in thermoelectric materials) affect $f$.[83] The second term of Equation (9) represents the effect of collisions (or, scattering) that prevents electrons from infinitely accelerating under external forces. The reason for Equation (9) = 0 is that electrons do not move outside in an open circuit at any time.

Under the relaxation time approximation,[84] *i.e.*

$$\left(\frac{\partial f}{\partial t}\right)_{collision} = \frac{f - f_0}{\tau} , \tag{10}$$

in which $\tau$ is the relaxation time of electrons, and $f_0$ is electron distribution at equilibrium as described in Equation (3).

The first term of BTE is

$$\left(\frac{\partial f}{\partial t}\right)_{field} = \frac{\partial f}{\partial t} + \frac{\partial f}{\partial \mathbf{x}}\frac{\partial \mathbf{x}}{\partial t} + \frac{\partial f}{\partial \mathbf{k}}\frac{\partial \mathbf{k}}{\partial t} . \tag{11}$$

Here, we want to derive the steady-state solution to the BTE, thus

$$\frac{\partial f}{\partial t} = 0 . \tag{12}$$

In addition, based on the momentum theory, we have

$$\hbar \partial \mathbf{k} = F \partial t , \tag{13}$$

namely

$$\frac{\partial \mathbf{k}}{\partial t} = \frac{F}{\hbar} , \tag{14}$$

in which, $\hbar$ is the reduced Planck constant, and $F$ is a generalized form of force. In the case of an electric field, $F$ equals to $e\varepsilon$.[85]

Through substituting Equations (12) and (14) into Equation (11), we have

$$\left(\frac{\partial f}{\partial t}\right)_{field} = \mathbf{v}\nabla_x f + \frac{F}{\hbar}\nabla_k f , \tag{15}$$

with the velocity vector (v),

$$\nabla_x f = \frac{\partial f}{\partial \mathbf{x}} = \frac{\partial f}{\partial E}\frac{\partial E}{\partial T}\frac{\partial T}{\partial \mathbf{x}} = \frac{\partial f}{\partial E}\left(\frac{E-u}{T}\right)\nabla_x T \text{ , and} \tag{16}$$

$$\nabla_k f = \frac{\partial f}{\partial \mathbf{k}} = \frac{\partial f}{\partial E}\frac{\partial E}{\partial \mathbf{k}} \overset{\frac{\partial E}{\partial \mathbf{k}}=\frac{\hbar^2 k}{m^*}}{=} \frac{\partial f}{\partial E}\frac{\hbar^2 k}{m^*} \text{ .} \tag{17}$$

Thus, Equation (15) becomes

$$\begin{aligned}\left(\frac{\partial f}{\partial t}\right)_{field} &= \mathbf{v}\frac{\partial f}{\partial E}\frac{E-u}{T}\nabla_x T + \frac{F}{\hbar}\frac{\partial f}{\partial E}\frac{\hbar^2 k}{m^*} = \frac{\partial f}{\partial E}\left(\mathbf{v}\frac{E-u}{T}\nabla_x T + \frac{F}{\hbar}\frac{\hbar^2 k}{m^*}\right) \\ &\overset{\mathbf{v}=\frac{\hbar k}{m^*}}{=} \frac{\partial f}{\partial E}\left(\mathbf{v}\frac{E-u}{T}\nabla_x T + \mathbf{v}F\right) = \frac{\partial f}{\partial E}\mathbf{v}\left(\frac{E-u}{T}\nabla_x T + F\right)\end{aligned} \text{ .} \tag{18}$$

Based on Equations (9), (10), and (18), the steady state solution to the BTE is

$$f - f_0 = -\mathbf{v}\tau\frac{\partial f}{\partial E}\left(\frac{E-u}{T}\nabla_x T + F\right) \overset{F=e\varepsilon,\, f\approx f_0}{=} -\mathbf{v}\tau\frac{\partial f_0}{\partial E}\left(\frac{E-u}{T}\nabla_x T + e\varepsilon\right) \tag{19}$$

### 3.3 Derivation of electronic transport coefficients

- Electrical conductivity

$$\begin{aligned}\sigma \equiv \left.\frac{J}{\varepsilon}\right|_{\nabla_x T=0} &= \frac{e}{\varepsilon}\int_{-\infty}^{+\infty}\mathbf{v}(E)g(E)\left[f(E)-f_0(E)\right]dE \\ &= \frac{e}{\varepsilon}\int_{-\infty}^{+\infty}\mathbf{v}(E)g(E)\mathbf{v}\tau\left(-\frac{\partial f_0}{\partial E}\right)\left(\frac{E-u}{T}\nabla_x T + e\varepsilon\right)dE \text{ .} \\ &= e^2\int_{-\infty}^{+\infty}\mathbf{v}^2(E)g(E)\tau\left(-\frac{\partial f_0}{\partial E}\right)dE\end{aligned} \tag{20}$$

- Carrier concentration ($n$)

Carrier concentration ($n$) is derived from the definition, namely

$$n = \int_0^\infty g(E)\, f_0(E)dE \text{ ,} \tag{21}$$

Then, drift mobility is

$$\mu = \frac{\sigma}{ne} \tag{22}$$

Practically, Hall effect is employed to characterize the electronic transport. Hall carrier concentration and Hall carrier mobility are given by

$$n_H = \frac{n}{A} \tag{23}$$

$$\mu_H = \mu A \tag{24}$$

in which $A$ is the Hall factor and can be expressed as

$$A=\frac{\dfrac{\int_0^{\infty}g(E)\tau^2(E)v^2(E)\left(-\dfrac{\partial f}{\partial E}\right)dE}{\int_0^{\infty}g(E)v^2(E)\left(-\dfrac{\partial f}{\partial E}\right)dE}}{\left(\dfrac{\int_0^{\infty}g(E)\tau(E)v^2(E)\left(-\dfrac{\partial f}{\partial E}\right)dE}{\int_0^{\infty}g(E)v^2(E)\left(-\dfrac{\partial f}{\partial E}\right)dE}\right)^2} \tag{25}$$

- Seebeck coefficient

The derivation of $S$ is based on $J = 0$. From Equations (4) and (19), we have

$$J=e\int_{-\infty}^{+\infty}v(E)g(E)\mathbf{v}\tau\left(-\frac{\partial f_0}{\partial E}\right)\left(\frac{E-u}{T}\nabla_x T+e\varepsilon\right)dE=0\ , \tag{26}$$

Thus,

$$\varepsilon=-\frac{\nabla_x T\int_{-\infty}^{+\infty}v(E)g(E)\mathbf{v}\tau\left(-\dfrac{\partial f_0}{\partial E}\right)\left(\dfrac{E-u}{T}\right)dE}{e\int_{-\infty}^{+\infty}v(E)g(E)\mathbf{v}\tau\left(-\dfrac{\partial f_0}{\partial E}\right)dE}\ , \tag{27}$$

So,

$$\begin{aligned}S\equiv\left.\frac{\varepsilon}{\nabla_x T}\right|_{J=0}&=-\frac{\nabla_x T\int_{-\infty}^{+\infty}v(E)g(E)\mathbf{v}\tau\left(-\dfrac{\partial f_0}{\partial E}\right)\left(\dfrac{E-u}{T}\right)dE}{\nabla_x Te\int_{-\infty}^{+\infty}v(E)g(E)\mathbf{v}\tau\left(-\dfrac{\partial f_0}{\partial E}\right)dE}\\&=-\frac{1}{eT}\left[\frac{\int_{-\infty}^{+\infty}\mathbf{v}^2(E)g(E)\tau E\left(-\dfrac{\partial f_0}{\partial E}\right)dE}{\int_{-\infty}^{+\infty}\mathbf{v}^2(E)g(E)\tau\left(-\dfrac{\partial f_0}{\partial E}\right)dE}-\mu\right]\end{aligned}\ . \tag{28}$$

- Electrical thermal conductivity and Lorenz number ($L$)

Based on Equations (4) and (5), we have

$$\begin{aligned}Q&=\int_{-\infty}^{+\infty}v(E)(E-u)g(E)\left[f(E)-f_0(E)\right]dE\\&=\int_{-\infty}^{+\infty}v(E)Eg(E)\left[f(E)-f_0(E)\right]dE-u\int_{-\infty}^{+\infty}v(E)g(E)\left[f(E)-f_0(E)\right]dE\\&=\int_{-\infty}^{+\infty}v(E)Eg(E)\left[f(E)-f_0(E)\right]dE-\frac{u}{e}e\int_{-\infty}^{+\infty}v(E)g(E)\left[f(E)-f_0(E)\right]dE\\&=\int_{-\infty}^{+\infty}v(E)Eg(E)\left[f(E)-f_0(E)\right]dE-\frac{u}{e}J\end{aligned}\ .$$

(29)

Since in the derivation of $\kappa_e$ $J = 0$,

$$Q=\int_{-\infty}^{+\infty}v(E)Eg(E)\left[f(E)-f_0(E)\right]dE\ . \tag{30}$$

Therefore,

$$\kappa_e \equiv -\frac{Q}{\nabla_x T}\bigg|_{J=0} = -\frac{1}{\nabla_x T}\int_{-\infty}^{+\infty} v(E)Eg(E)\left[f(E)-f_0(E)\right]dE$$
$$= -\frac{1}{T}\int_{-\infty}^{+\infty}\mathbf{v}^2(E)Eg(E)\tau\left(-\frac{\partial f_0}{\partial E}\right)(E-u)dE - \frac{e\varepsilon}{\nabla_x T}\int_{-\infty}^{+\infty}\mathbf{v}^2(E)Eg(E)\tau\left(-\frac{\partial f_0}{\partial E}\right)dE \quad (31)$$

Substituting Equation (27) into above equation, we have

$$\kappa_e = -\frac{1}{T}\int_{-\infty}^{+\infty}\mathbf{v}^2(E)Eg(E)\tau\left(-\frac{\partial f_0}{\partial E}\right)(E-u)dE + \frac{\int_{-\infty}^{+\infty}\mathbf{v}^2(E)g(E)\tau\left(-\frac{\partial f_0}{\partial E}\right)\left(\frac{E-u}{T}\right)dE}{\int_{-\infty}^{+\infty}\mathbf{v}^2(E)g(E)\tau\left(-\frac{\partial f_0}{\partial E}\right)dE}\int_{-\infty}^{+\infty}\mathbf{v}^2(E)Eg(E)\tau\left(-\frac{\partial f_0}{\partial E}\right)dE$$
$$= -\frac{1}{T}\int_{-\infty}^{+\infty}\mathbf{v}^2(E)E^2g(E)\tau\left(-\frac{\partial f_0}{\partial E}\right)dE + \frac{\left(\int_{-\infty}^{+\infty}\mathbf{v}^2(E)g(E)\tau E\left(-\frac{\partial f_0}{\partial E}\right)dE\right)^2}{T\int_{-\infty}^{+\infty}\mathbf{v}^2(E)g(E)\tau\left(-\frac{\partial f_0}{\partial E}\right)dE} \quad (32)$$

According to Wiedemann–Franz law,[86] Lorenz number is

$$L \equiv \frac{\kappa_e}{\sigma T}\,. \quad (33)$$

Substituting Equations (20) and (32) into above Equation (33), we have

$$L = -\frac{1}{T^2e^2}\left(\frac{\int_{-\infty}^{+\infty}\mathbf{v}^2(E)E^2g(E)\tau\left(-\frac{\partial f_0}{\partial E}\right)dE}{\int_{-\infty}^{+\infty}\mathbf{v}^2(E)g(E)\tau\left(-\frac{\partial f_0}{\partial E}\right)dE} - \left(\frac{\int_{-\infty}^{+\infty}\mathbf{v}^2(E)g(E)\tau E\left(-\frac{\partial f_0}{\partial E}\right)dE}{\int_{-\infty}^{+\infty}\mathbf{v}^2(E)g(E)\tau\left(-\frac{\partial f_0}{\partial E}\right)dE}\right)^2\right). \quad (34)$$

So far, we derived $\sigma$, $S$, $\kappa_e$, and $L$, expressed by Equations (20), (28), (32), and (34), respectively.

### 3.4 Kane band model

In the non-parabolic band with energy dispersion described by Kane relation, the DOS can be expressed as[87]

$$g(\mathrm{E}) = \frac{\sqrt{2}N_v m_b^{*3/2}}{\pi^2\hbar^3}E^{1/2}\left(1+\frac{E}{E_g}\right)^{1/2}\left(1+2\frac{E}{E_g}\right) \quad (35)$$

Under the assumption of acoustic phonons dominating carrier scattering, carrier relaxation time is[88,89]

$$\tau(\varepsilon)=\frac{2\pi\hbar^4 C_l}{(k_B T)^2 (2m_b^*)^{3/2} E_{def}^2} E^{-1/2}\left(1+\frac{E}{E_g}\right)^{-1/2}\left(1+2\frac{E}{E_g}\right)^{-1}\left[1-\frac{8\frac{E}{E_g}\left(1+\frac{E}{E_g}\right)}{3\left(1+2\frac{E}{E_g}\right)^2}\right]^{-1} \quad (36)$$

By substituting $g(E)$ and $\tau(E)$ into the derived thermoelectric properties, we can have the single Kane band model[90]

$$S=\frac{k_B}{e}\left[\frac{F_{1,-2}^1(\eta,\beta)}{F_{1,-2}^0(\eta,\beta)}-\eta\right] \quad (37)$$

$$\sigma=\frac{2e^2 N_v \hbar C_l}{\pi m_I^* E_{def}^2} F_{1,2}^0(\eta,\beta) \quad (38)$$

$$L=\left(\frac{k_B}{e}\right)^2\left[\frac{F_{1,-2}^2(\eta,\beta)}{F_{1,-2}^0(\eta,\beta)}-\left(\frac{F_{1,-2}^1(\eta,\beta)}{F_{1,-2}^0(\eta,\beta)}\right)^2\right] \quad (39)$$

Hall carrier concentration

$$n_H=\frac{N_v\left(2m_b^* k_B T\right)^{\frac{3}{2}}}{3\pi^2\hbar^3}\frac{F_{3/2,0}^0(\eta,\beta)}{A} \quad (40)$$

Hall carrier mobility

$$\mu_H=A\frac{2\pi\hbar^4 e C_l}{m_I^*\left(2m_b^* k_B T\right)^{3/2} E_{def}^2}\frac{3F_{1,-2}^0(\eta,\beta)}{F_{3/2,0}^0(\eta,\beta)} \quad (41)$$

Hall factor

$$A=\frac{3K(K+2)}{(2K+1)^2}\frac{F_{1/2,-4}^0 F_{3/2,0}^0}{\left(F_{1,-2}^0\right)^2} \quad (42)$$

Generalized Fermi integration

$$F_{m,k}^n(\eta,\beta)=\int_0^\infty\left[-\frac{\partial f(\eta)}{\partial\varepsilon}\right]\varepsilon^n\left(\varepsilon+\beta\varepsilon^2\right)^m\left[(1+2\beta\varepsilon)^2+2\right]^{\frac{k}{2}} d\varepsilon \quad (43)$$

where $\eta$ is the reduced Fermi level (for electrons, $\eta_e=\frac{E_f-E_c}{k_B T}$ with $E_c$ denoting the conduction band edge; for holes, $\eta_h=\frac{E_v-E_f}{k_B T}=\frac{E_c-E_g-E_f}{k_B T}=-1/\beta-\eta_e$ with $E_v$ denoting the valance band edge), $\beta$ is reciprocal of reduced band gap (*i.e.* $\beta=\frac{k_B T}{E_g}$ ), $k_B$ is the Boltzmann constant, $\hbar$ is the reduced Planck constant, $N_v$ is the band degeneracy, $K$ is the ratio of longitudinal ($m_{||}^*$) and transverse ($m_\perp^*$) effective mass, $C_l$ is the combination of elastic constants, $m_b^*$ is the band effective mass, $m_I^*$ is the inertial effective mass, $e$ is free electron charge, $m_0$ is the free electron mass, and $E_{def}$ is the deformation potential, respectively. The relations of $m_b^*$, $m_I^*$, and density of state (DOS) effective mass ($m_d^*$) are expressed as

$$m_b^* = N^{-\frac{2}{3}} m_d^* \tag{44}$$

and

$$m_I^* = N^{-\frac{2}{3}} m_d^* \frac{3K^{\frac{2}{3}}}{2K+1} \tag{45}$$

In some cases with notable bipolar conduction, we should consider both valance band and conduction band.[91,92]

Total Seebeck coefficient

$$S_{tot} = \frac{S_{CB}\sigma_{CB} + S_{VB}\sigma_{VB}}{\sigma_{CB} + \sigma_{VB}} \tag{46}$$

Total electrical conductivity

$$\sigma_{tot} = \sigma_{CB} + \sigma_{VB} \tag{47}$$

Total Hall coefficient

$$R_{Htot} = \frac{R_{HCB}\sigma_{CB}^2 + R_{HVB}\sigma_{VB}^2}{\left(\sigma_{CB} + \sigma_{VB}\right)^2} \tag{48}$$

Bipolar thermal conductivity

$$\kappa_{bi} = \frac{\sigma_{CB}\sigma_{VB}}{\sigma_{CB} + \sigma_{VB}} \left(S_{CB} - S_{VB}\right)^2 T \tag{49}$$

Total value of $L$

$$L_{tot} = \frac{L_{CB}\sigma_{CB} + L_{VB}\sigma_{VB}}{\sigma_{CB} + \sigma_{VB}} \tag{50}$$

In the above equations, the components contributed by conduction band (CB) and valence band (VB) are presented by the corresponding subscripts. It should be mentioned that $S$ and $R_H$ are positive for *p*-type but negative for *n*-type in those equations.

## 4 Modeling studies of the parameters affecting thermoelectric properties of $Bi_2Te_3$

Based on above equations, we calculated the electronic transport coefficients as a function of $\eta$ for $Bi_2Te_3$. Fig. 3a shows the calculated $\sigma$ as a function of $\eta$ at 300 K with the inset illustrating the variation of $\eta$ in the band structure. Note that we used $E_g$ = 0.15 eV, and set the VB maximum as 0, which means the CB minimum is at 0.15 eV in energy. Therefore, at 300 K, $\eta$ = -5.8 corresponding to the CB minimum. Moreover, considering the identical band effective mass for CB and VB, the middle point of band gap region is at $\eta$ = ~ -3. As a consequence, $\sigma$ for *n*-type (left side of -3) and *p*-type (right side of -3) is identical, and the total $\sigma$ (blue curve) is generally superimposed with the CB component (red curve) for *n*-type and VB component (green curve) for *p*-type, respectively. Furthermore, in the whole studied $\eta$ range, $\sigma$ increases with enlarging $|\eta|$ for either *n*-type or *p*-type. The observed characteristics of symmetry (*i.e.* the calculated curve as a function of $\eta$ is symmetrical relative to the middle of band gap region.) and superimposition (*i.e.* the total value is respectively superimposed with the CB/VB component for *n*/*p*-type situation.) also appear in the other calculated thermoelectric properties.

Fig. 3b shows the calculated $S$ as a function of $\eta$ at 300 K. The characteristics of symmetry and superimposition are also observed in $|S|$. When -5.8 < $\eta$ < 0, the total $S$ is different from each component contributed by either CB or VB, which reveals *n*/*p*- type transition, *i.e.* the bipolar conduction. The peak value of $|S|$ depends on the onset of bipolar conduction, and $|S|$ decreases with increasing $|\eta|$ for either *n*-type or *p*-type. Based on the calculated $S$ and $\sigma$, the plots of $S^2\sigma$ as a function of $\eta$ at 300 K were obtained, as shown in Fig. 3c. Likewise, $S^2\sigma$ also shows symmetrical and superimposing characteristics. Because of the bipolar conduction, the positions of total $S^2\sigma$ peaks are slightly different from the CB/VB components. If the $\eta$ corresponding to the peak of $S^2\sigma$ is defined as the optimized $\eta$ ($\eta^{opt}$), $\eta^{opt}$ for is 0.25 and -6.05 for *p*-type and *n*-type, respectively.

Fig. 3d shows the plots of $L$ as a function of $\eta$ at 300 K. We can observe the superimposing and symmetrical features, and $L$ increases with increasing $|\eta|$. Based on the calculated $\sigma$ and $L$, we calculated $\kappa_e$, shown in Fig. 3e, from which $\kappa_e$ increases with increasing $|\eta|$. Fig. 3f shows the calculated $\kappa_{bi}$. As can be seen, $\kappa_{bi}$ is prominent in the band gap region and reaches the maximum value in the middle of band gap region. Moreover, $\kappa_{bi}$ decreases with increasing $|\eta|$, because large $|\eta|$ makes it more difficult to activate minor carriers.

Based on above discussions, $S$ and $\sigma$ are inversely related to $\eta$, resulting in a peak for $S^2\sigma$. Bipolar conduction has further limited the maximum values of $S$ for *n*/*p*-type materials, therefore, the possible maximum values for $S^2\sigma$ is even smaller by considering both CB and VB. In addition to the electrical energy, the charge carriers can also transport thermal energy in the form of $\kappa_e$ and $\kappa_{bi}$, which are anticipated to reduce the ultimate $zT$.

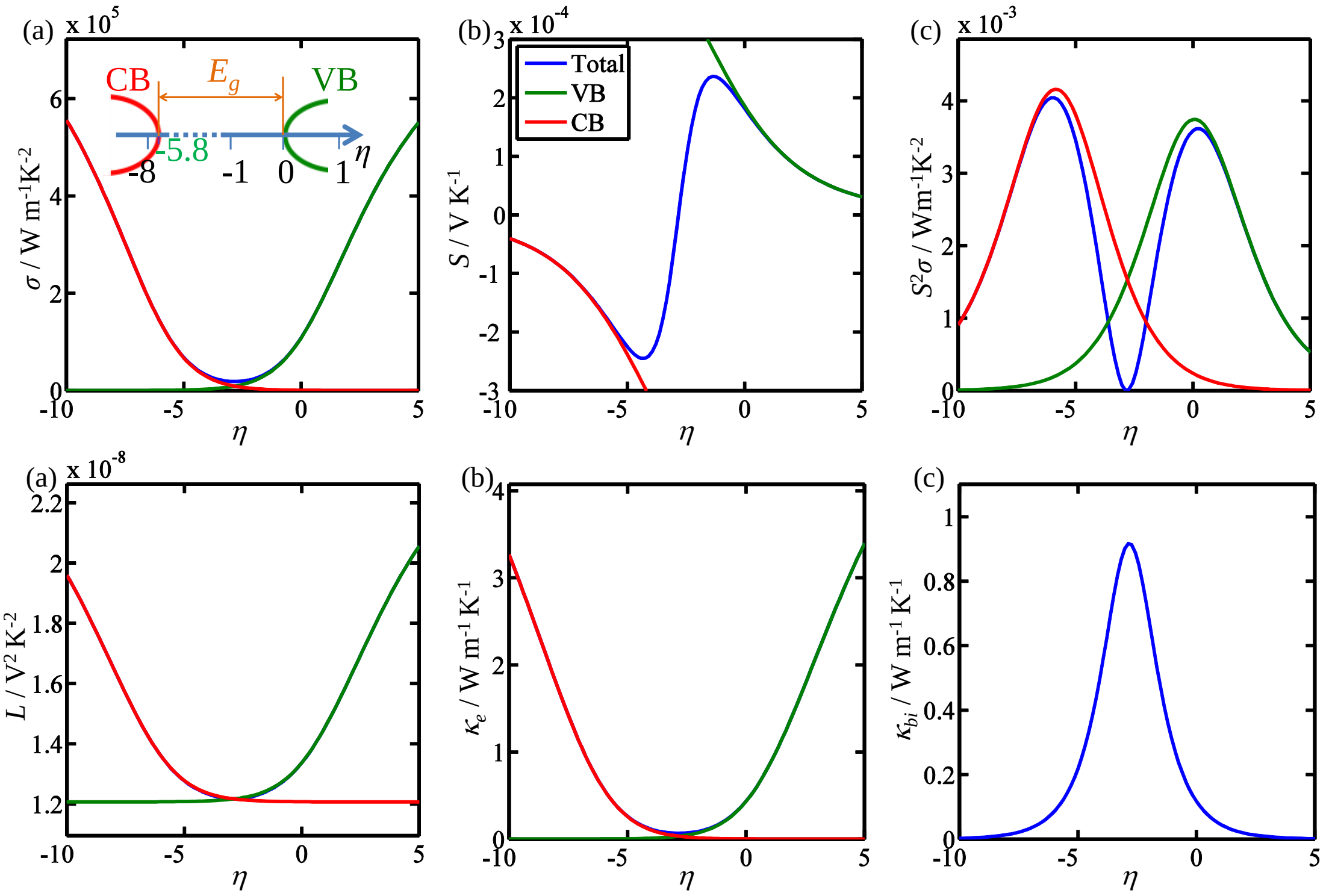

Fig. 3. (color online) Calculated (a) $S$, (b) $\sigma$, (c) $S^2\sigma$, (d) $L$, (e) $\kappa_e$, and (f) $\kappa_e$ as a function of $\eta$ at 300 K with the blue curve representing the total values, the purple curve representing the CB component, green curve representing the VB component.

Based on the Kane band model, we can see $S$ is only related to $\eta$ and $\beta$ (= $k_BT/E_g$). Apart from $\eta$ and $\beta$, $n_H$ and $\mu_H$ are determined by $m_d^*$, $m_I^*$, and $E_{def}$. Therefore, $\sigma$, $\kappa_e$ and $\kappa_{bi}$ are also related to these parameters. Through band engineering, we can tune these parameters to tailor the electronic transport. Here, we quantitatively predict the effects of $m_d^*$, $E_{def}$, and $E_g$ on the electronic transport properties. Note that from the Kane model equations, all thermoelectric properties are direct as a function of $\eta$. Here, we want to examine the variations of these thermoelectric properties with $n_H$.

Fig. 4a shows the determined $\eta$ with $n_H$ ranging from $10^{18}$ to $10^{21}$ cm$^{-3}$ for evenly selected seven $m_d^*$ values from 0.2 $m_0$ to 2 $m_0$. With increasing $n_H$, $\eta$ monotonically increases, and with increasing $m_d^*$, $\eta$ decreases for a given $n_H$, which results in the $E_f$ corresponding to high $n_H$ still resides in the band gap region for large $m_d^*$. Thus, larger $m_d^*$ would unfavorably lead to a strong bipolar conduction. Based on the determined $\eta$, Fig. 4b – d show the effects of $m_d^*$ on $n_H$ dependent $S$, $\mu_H$, and $S^2\sigma$, respectively. With increasing $n_H$, both $S$, and $\mu_H$ increase and then decrease. Moreover, we observed that the calculated variation of $S$ and $\mu_H$ at low $n_H$ is different from the monotonic decreasing trends in both $S$ and $\mu_H$ with increasing $n_H$ calculated by the single band model (refer to the bold green lines in Fig. 4b and c). The observed difference suggests that, at low $n_H$ region (band gap region), both $S$ and $\mu_H$ calculated by the two bands model are lower than those from single band model, and the reason is ascribed to the bipolar conduction. Moreover, with increasing $m_d^*$, $S$ increases, while $\mu_H$ decreases, and $\mu_H$ is more sensitive to the variation of $m_d^*$. As a consequence, $S^2\sigma$ decreases with increasing $m_d^*$, as shown in Fig. 4d. In addition, optimal $n_H$ ($n_H^{opt}$) corresponding to the peak $S^2\sigma$ shifts to a higher value with increasing $m_d^*$.

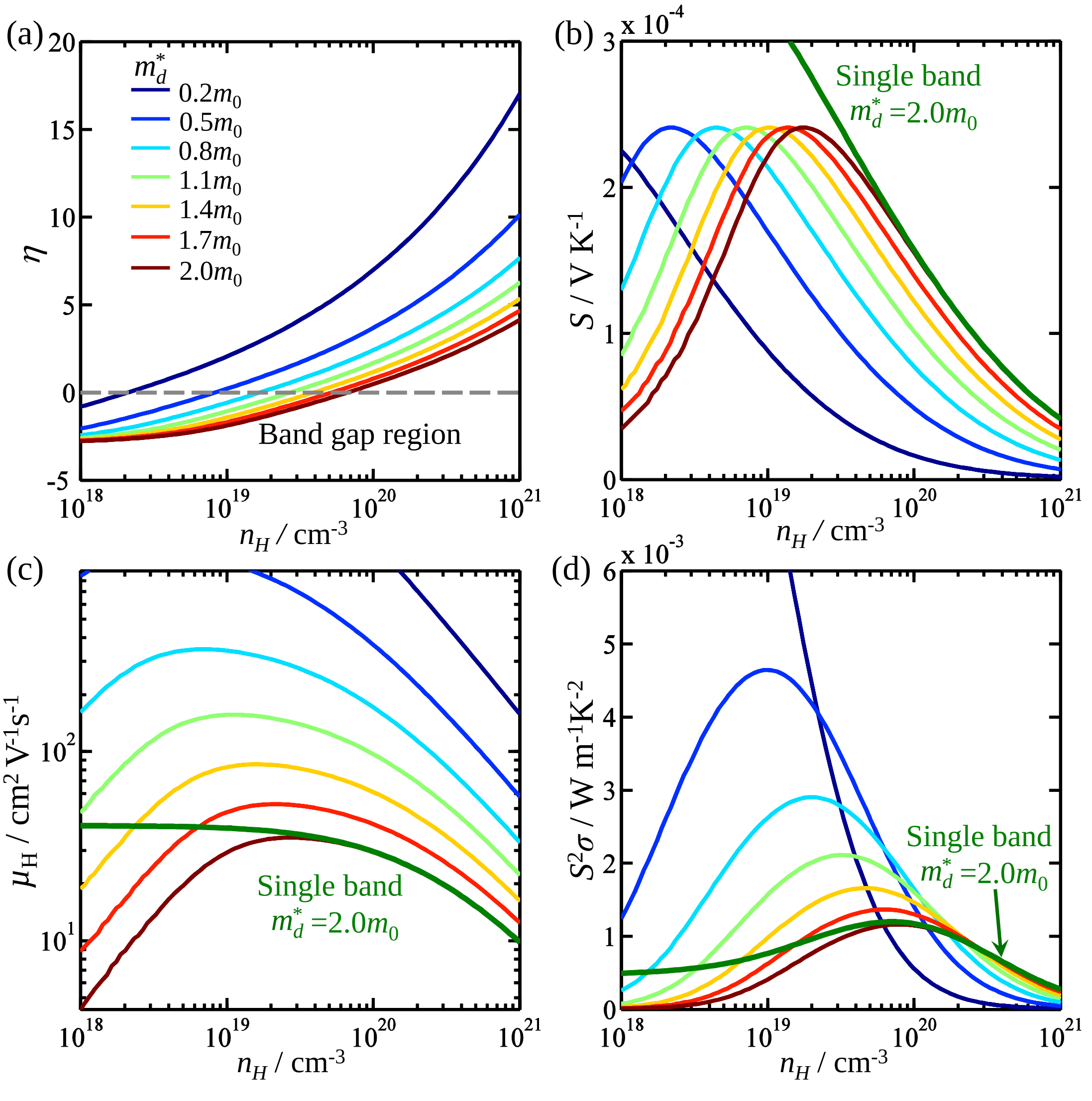

Fig. 4. (color online) (a) Determined $\eta$ with $n_H$ ranging from $10^{18}$ to $10^{21}$ cm$^{-3}$ for evenly selected seven $m_d^*$ values from 0.2 $m_0$ to 2 $m_0$. Correspondingly calculated (b) $S$, (c) $\mu_H$,

and (d) $S^2\sigma$ as a function of $n_H$ for different $m_d^*$ values. The bold green lines in (b) and (c) are calculated using single band model with $m_d^*$ = 2 $m_0$ for $S$ and $\mu_H$, respectively.

Fig. 5 shows the effects of $E_g$ on thermoelectric properties. Firstly, with increasing $n_H$ from $10^{18}$ to $10^{21}$ cm$^{-3}$, the determined $\eta$ for evenly selected seven $E_g$ values from 0.1 to 0.5 eV is exhibited in Fig. 5a. As can be seen, $|\eta|$ tends to increases with increasing $E_g$ for a given $n_H$. Fig. 5b presents the plots of $S$ as a function of $n_H$ for different $E_g$ values. Larger $E_g$ produces larger $S$ peak and shifts the peak of $S$ to low $n_H$, which quantitatively verifies the effectiveness of $E_g$ on suppressing bipolar conduction. The suppressed bipolar conduction due to large $E_g$ is also demonstrated in the variation of $\mu_H$ (see Fig. 5c). Fig. 5d shows the variations of $S^2\sigma$ with $E_g$, from which enlarging $E_g$ can greatly enhance peak $S^2\sigma$ but does not change the $n_H^{opt}$ significantly.

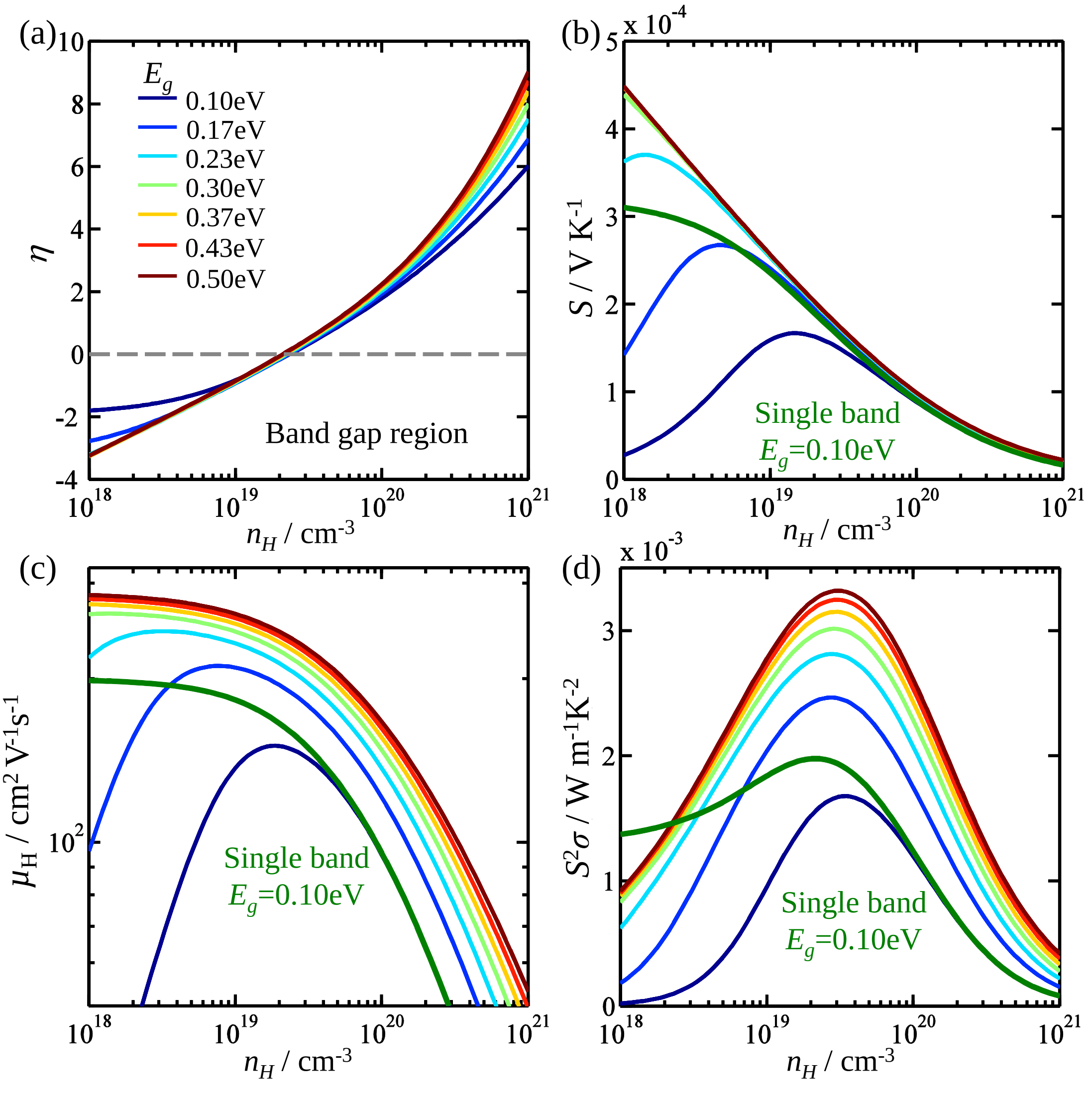

Fig. 5. (color online) (a) Determined $\eta$ with $n_H$ ranging from $10^{18}$ to $10^{21}$ cm$^{-3}$ for evenly selected seven $E_g$ values from 0.1 to 0.5 eV. Correspondingly calculated (b) $S$, (c) $\mu_H$, and (d) $S^2\sigma$ as a function of $n_H$ for different $E_g$ values.

Lastly, we studied the impacts of $E_{def}$ on electronic transport coefficients. Because $E_{def}$ characterizes the strength of phonon scattering on free charge carriers,[93,94] we anticipate that $E_{def}$ would not affect $\eta$ and bipolar conduction. This has been confirmed by the determined $\eta$ and the correspondingly calculated $S$, as shown in Fig. 6a and b, respectively. However, small $E_{def}$ can greatly increase $\mu_H$ (refer to Fig. 6c). As a consequence, small $E_{def}$ leads to high $S^2\sigma$, which can be seen from Fig. 6d. Interestingly, $E_{def}$ does not change $n_H^{opt}$ .

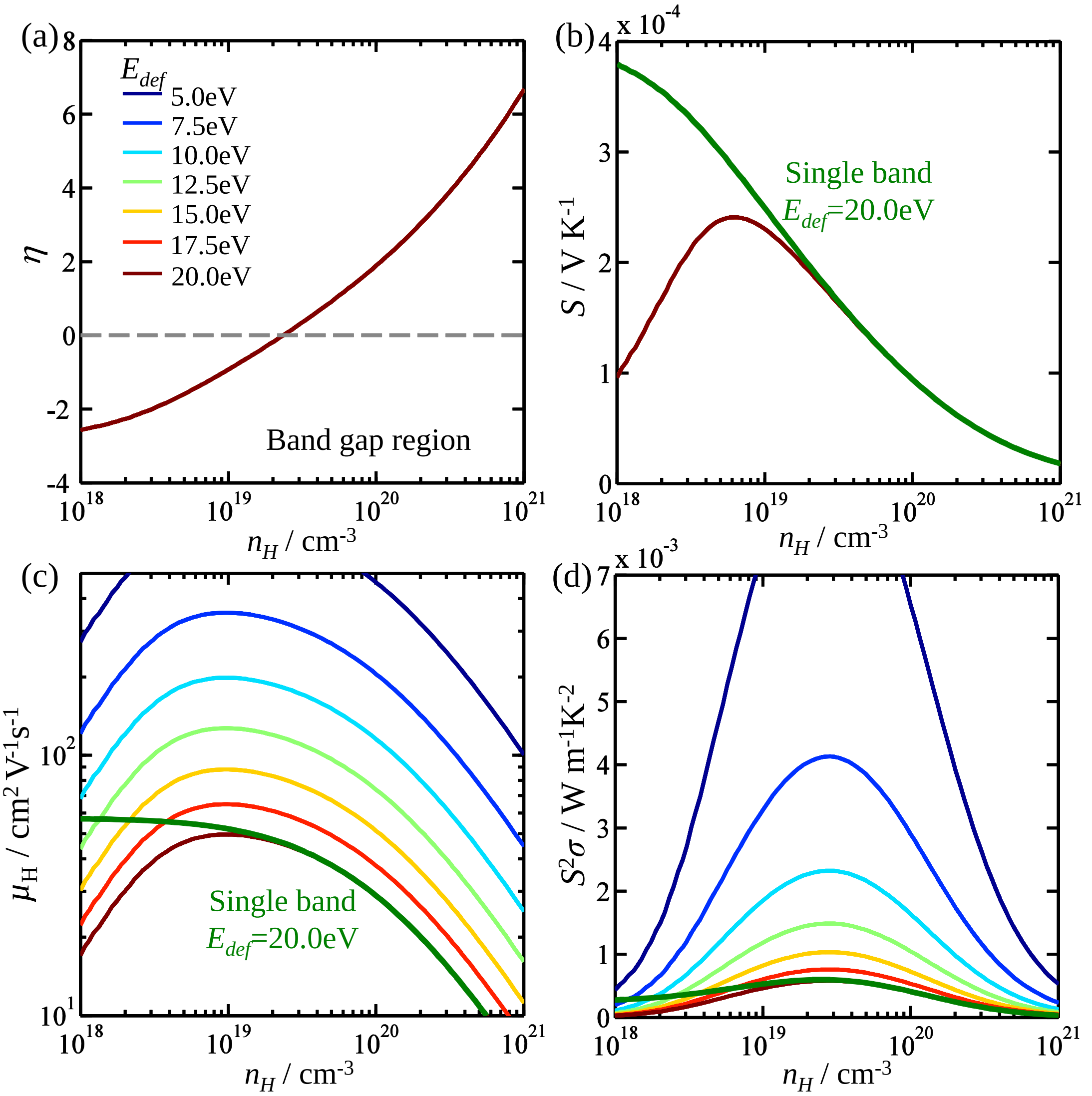

Fig. 6. (color online) (a) Determined $\eta$ with $n_H$ ranging from $10^{18}$ to $10^{21}$ cm$^{-3}$ for evenly selected seven $E_{def}$ values from 5 to 20 eV. Correspondingly calculated (b) $S$, (c) $\mu_H$, and (d) $S^2\sigma$ as a function of $n_H$ for different $E_{def}$ values. The bold green lines in (b) and (c) are calculated using SKB model with $E_{def}$ = 8 eV for $S$ and $\mu_H$, respectively.

It is well-documented that bipolar conduction is detrimental to the thermoelectric performance by generating $\kappa_{bi}$ and reducing $S$. Here, we emphasized the significance of suppressing bipolar conduction by enlarging $E_g$. Fig. 7a and b show the calculated $\kappa_{bi}$ and $S$ as a function of $n_H$ for evenly selected seven $E_g$ values from 0.1 to 0.5 eV. Because bipolar conduction is more notable at high temperature, the calculation covers the results at 300 K, 400 K, and 500 K. As can be seen, at high temperature, $\kappa_{bi}$ is large but $S$ is small, and enlarging $E_g$ can greatly reduce $\kappa_{bi}$ but increases $S$.

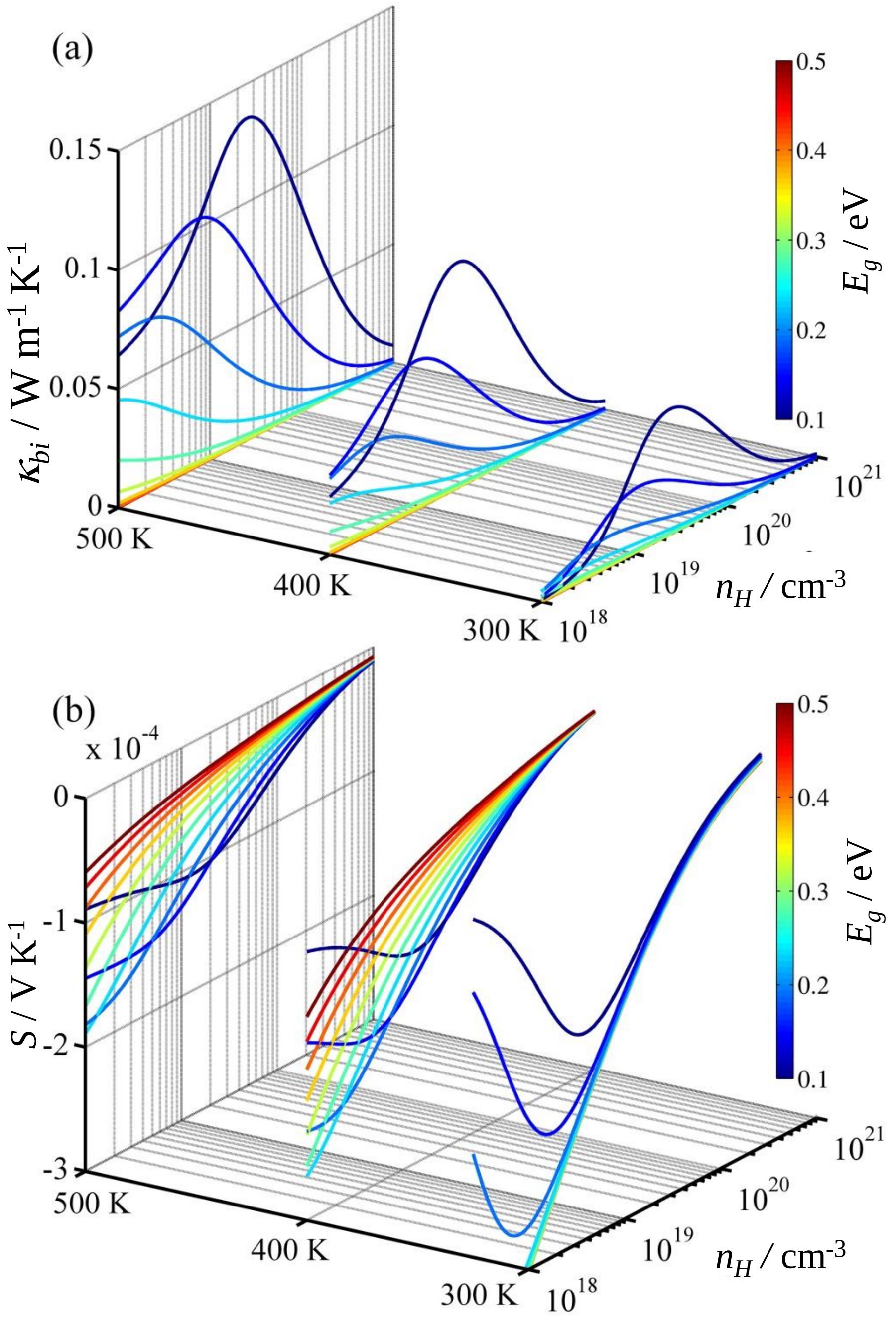

Fig. 7. (color online) Calculated (a) $\kappa_{bi}$, and (b) $S$ as a function of $n_H$ at 300 K, 400 and 500 K for evenly selected seven $E_g$ values from 0.1 to 0.5 eV.

Based on above discussions, we can see that $m_d^*$, $E_g$, and $E_{def}$ can significantly affect $S^2\sigma$, but only $m_d^*$ is able to change the $n_H^{opt}$. To determine $n_H^{opt}$, we calculated $S^2\sigma$ as functions of $n_H$ and $m_d^*$, shown in Fig. 8a. Based on this, the $n_H^{opt}$ values corresponding to different $m_d^*$ are plot as a white curve in Fig. 8b, in which the background is the contour plot of $S^2\sigma$ as functions of $n_H$ and $m_d^*$. As can be seen, the variation of contour plot follows the white curve. To enhance thermoelectric performance, we want to ensure the experimental $n_H$ value close to $n_H^{opt}$.

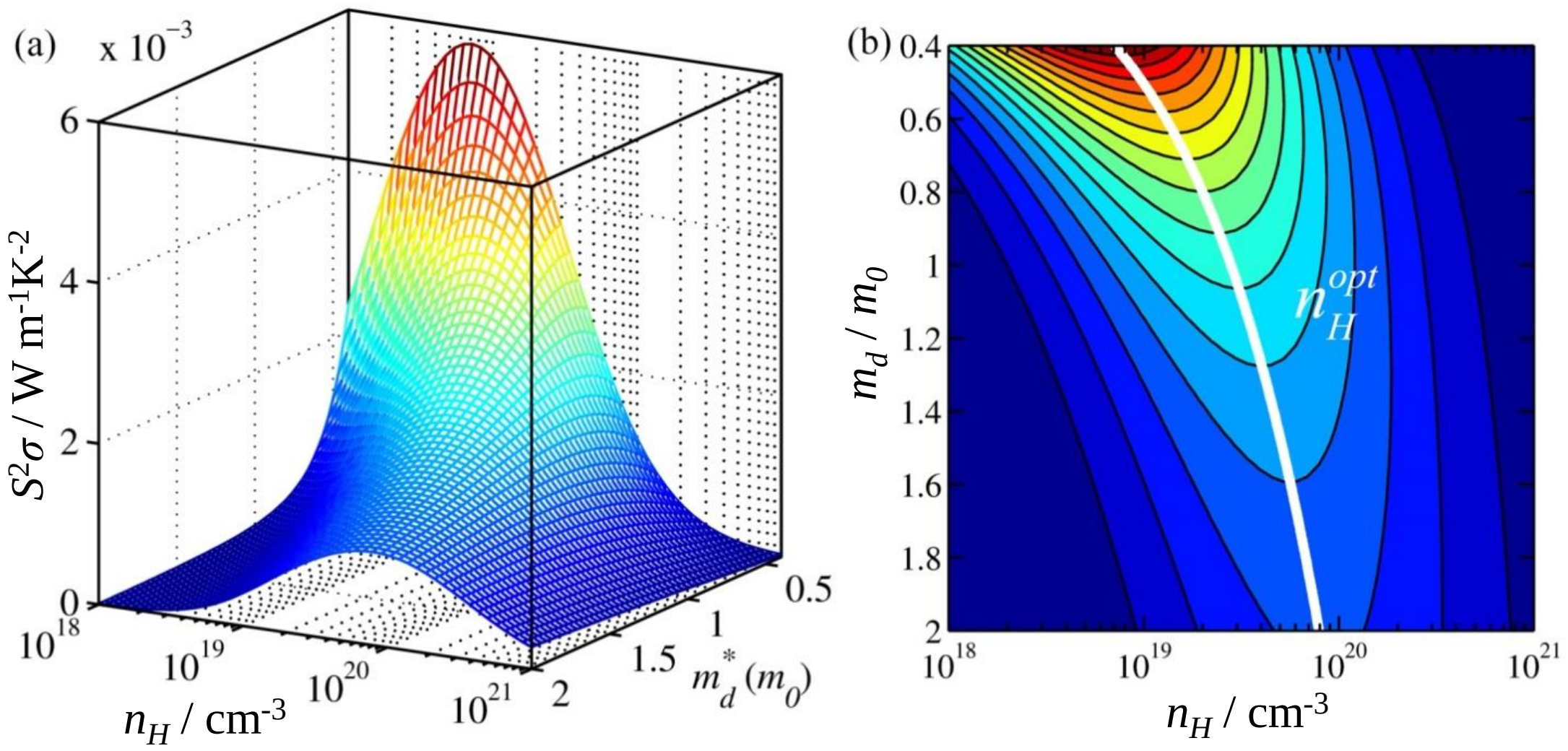

Fig. 8. (color online) (a) Calculated $S^2\sigma$ as functions of $n_H$ and $m_d^*$ , and (b) the corresponding contour map with the white curve indicating the $m_d^*$ dependent $n_H^{opt}$ .

## 5 $Bi_2Te_3$-based thermoelectric materials

$Bi_2Te_3$-based materials share the same rhombohedral crystal structure of the space group $R\bar{3}m$ (see Fig. 9a). This category consists of five-atom layers arranged along the $c$-axis, known as quintuple layers. The coupling is strong between two atomic layers, but is much weaker within one quintuple layer, predominantly of the van der Waals type. Lattice parameters and band gap ($E_g$) of these layered materials are shown in Table 1. The electronic band structures of $Bi_2Te_3$ are shown in Fig. 9c with the selected high symmetry $k$ points shown in Fig. 9b. As can be seen, both the highest valence band and lowest conduction band have six valleys. Besides these two bands, the second conduction and valence band with energy separations of 30 meV and 20 meV, respectively.[95]

Pnictogen (Bi and Sb) and chalcogenides (Te and Se) materials have been preferably studied for room-temperature thermoelectric applications.[70] These materials share the same rhombohedral crystal structure of the space group $R\bar{3}m$ (see Fig. 9). This category consists of five-atom layers arranged along the $c$-axis, known as quintuple layers. The coupling is strong between two atomic layers, but is much weaker within one quintuple layer, predominantly of the van der Waals type. Lattice parameters and band gap ($E_g$) of these layered materials are shown in Table 1.

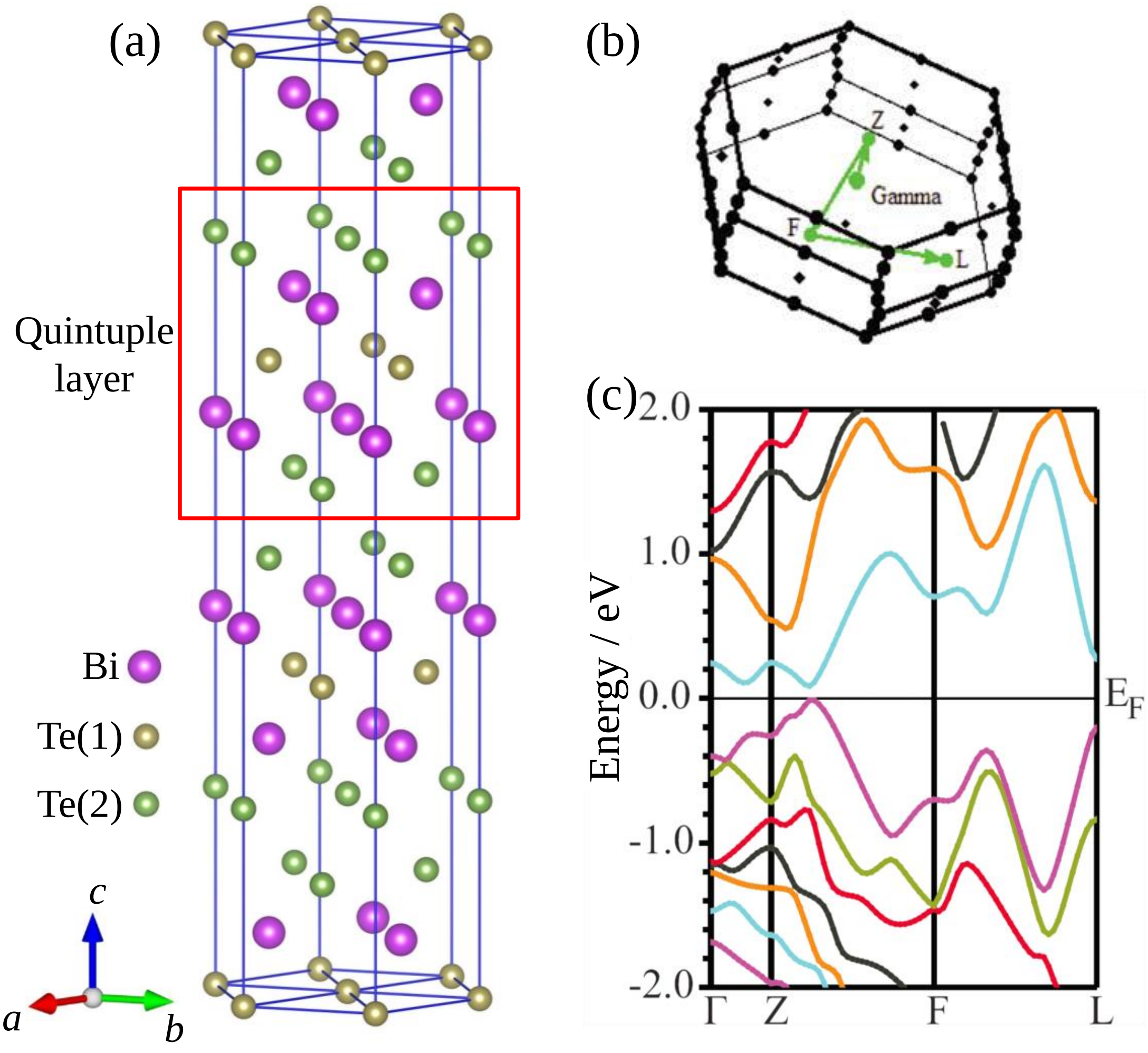

Fig. 9. (color online) (a) crystal structure, (b) first Brillouin zone, and (c) band structure. Reproduced from Ref. [96].

Table 1. Physical properties of $N_2M_3$ (N: Bi, Sb; M: Te, Se).[97,98]

| | $Bi_2Te_3$ | $Bi_2Se_3$ | $Sb_2Te_3$ |
|---|---|---|---|
| Structure | Hexagonal | Hexagonal | Hexagonal |
| $a$ / Å | 4.38 | 4.14 | 4.26 |
| $c$ / Å | 30.48 | 28.64 | 30.45 |
| Unit layer / Å | 10.16 | 9.55 | 10.15 |
| Band Gap / eV | 0.15 | 0.3 | 0.22 |

Based on above discussions, $Bi_2Te_3$ families crystalize in layered structures and consist of heavy atoms, which can potentially ensure low $\kappa$. Moreover, the narrow $E_g$ can secure a high $\sigma$, and the large band degeneracy is also beneficial to produce a high $S^2\sigma$. Because of these advantages of $Bi_2Te_3$ for thermoelectric applications, great efforts have been dedicated to enhancing the thermoelectric efficiencies. Table 2 summarizes the reported thermoelectric properties for both $n$-type $Bi_2Te_3$ and $p$-type $Sb_2Te_3$ based thermoelectric materials. As can be seen, $zT$ of 1.2 has been achieved in $n$-type $Bi_2Te_3$,[99-102] and $zT$ of 1.56 has been obtained in $p$-type $Sb_2Te_3$.[103] The big difference between the $n$-type and the $p$-type materials is mainly due to the much lower $S^2\sigma$ in $Bi_2Te_3$, although the obtained lowest $\kappa$ of *n*-type $Bi_2Te_3$ is even smaller than that of *p*-type $Sb_2Te_3$.

Table 2. Thermoelectric properties of the $Bi_2Te_3$-based materials prepared by different methods.

| Material | Type | $S^2\sigma$ / $10^{-4}$ $Wm^{-1}K^{-2}$ | $\kappa$ ($\kappa_l$) / $Wm^{-1}K^{-1}$ | $zT$ | $T$ / K | Method* |
|---|---|---|---|---|---|---|
| Nanostructuring | | | | | | |
| $Bi_2Te_3$[104] | *n* | 11.9 | 0.46(0.25) | 0.91 | 350 | MSS+CP |
| $Bi_{0.5}Sb_{1.5}Te_3$[104] | *P* | 14.9 | 0.45(0.25) | 1.2 | 363 | MSS+CP |
| $Bi_2Te_{2.7}Se_{0.3}$[105] | *n* | 11 | 0.6 | 0.55 | 300 | SG+SPS |
| $Bi_2Te_{2.7}Se_{0.3}$[106] | *n* | 11 | 0.6 | 0.54 | 300 | SG+SPS |
| $Bi_{0.5}Sb_{1.5}Te_3$[38] | *p* | 28 | 0.7(0.4) | 1.2 | 320 | MSS+SPS |
| $Bi_{0.4}Sb_{1.7}Te_{3.0}$[107] | *p* | 9 | 0.35(0.16) | 0.9 | 413 | SG+SPS |
| $(Bi_2Te_3)_{0.85}$ $(Bi_2Se_3)_{0.15}$[108] | *n* | 12 | 0.68(0.45) | 0.71 | 480 | SG+SPS |
| $(Bi_2Te_3)_{0.8}$ $(Bi_2Se_3)_{0.2}$[109] | *n* | 9 | 0.53(0.38) | 0.71 | 450 | SG+SPS |
| $Bi_{0.5}Sb_{1.5}Te_3$[110] | *p* | 24 | 0.66(0.3) | 1.13 | 360 | MSS+SPS |
| $Bi_2Te_3$[111] | *n* | 15.4 | 1.1 | 0.66 | 470 | SG+HP |
| $Bi_2Te_3$-Te[112] | *n* | 18.7 | 1.22(0.45) | 0.6 | 390 | SG+HP |
| $Bi_2Te_3$[35] | *n* | 6.9 | 0.45(0.28) | 0.62 | 400 | SG+SPS |
| $Bi_2Se_3$[113] | *n* | 4.4 | 0.42 | 0.35 | 400 | LIE+HP |
| $Bi_2Se_3$[114] | *n* | 4.7 | 0.41(0.3) | 0.48 | 425 | MSS+SPS |
| S doped $Sb_2Te_3$[115] | *p* | 20.0 | 0.7(0.35) | 0.95 | 423 | MSS+CP |
| Bulk materials | | | | | | |
| $Bi_2Te_3$[116] | *p* | 20 | 0.78(0.36) | 1.03 | 403 | HPS |
| $Bi_2Se_3$[117] | *n* | 6 | 0.97(0.4) | 0.37 | 560 | HPS |
| $Bi_{0.5}Sb_{1.5}Te_3$[70] | *p* | 44 | 1 | 1.4 | 373 | BM+HP |
| $Bi_{0.5}Sb_{1.5}Te_3$[118] | *p* | 43 | 1 | 1.3 | 373 | BMA+HP |
| $Bi_2Te_{2.7}Se_{0.3}$ [119] | *n* | 26 | 1.08 | 1.04 | 400 | BMA+HP |
| $Bi_2Te_3$[102] | *n* | 33 | 1.1 | 1.2 | 425 | BMA+HP |
| $Bi_{0.5}Sb_{1.5}Te_3$[120] | *p* | 38 | 0.85(0.48) | 1.4 | 300 | MA+HP |
| $Bi_2Te_{2.79}Se_{0.21}$[99] | *n* | 42 | 0.8(0.56) | 1.2 | 357 | HP |
| $Bi_2Te_{2.3}Se_{0.7}$[100] | *n* | 28 | 1.1(0.4) | 1.2 | 445 | MA+BM+ HP |
| $Bi_{0.3}Sb_{1.7}Te_3$[100] | *p* | | | 1.3 | 380 | MA+BM+ HP |
| $Bi_2Te_{2.925}Se_{0.075}$[121] | *n* | 47 | 1.65(1.27) | 0.85 (*a*-*b*) | 293 | THM |
| $Bi_{0.52}Sb_{1.48}Te_3$[103] | *p* | 35 | 0.65(0.25) | 1.56 | 300 | MS |
| $Cu_{0.01}Bi_2Te_{2.7}$ $Se_{0.3}$[122] | *n* | 31.2 | 1.1 | 1.06 | 373 | BM+HP |
| $Bi_2Te_{2.7}Se_{0.3}$[101] | *n* | 53.8 | 1.87 | 1.18 | 410 | BS |
| $Bi_2(Te_{1-x}Se_x)_3$ -I(0.08%)[123] | *n* | 55 | 1.5(0.9) | 1.1 | 340 | ZM |
| $Bi_2(Te_{0.5}Se_{0.5})_3$ -I 0.1%[123] | *n* | 25 | 1.42(0.45) | 0.85 | 570 | ZM |

* The abbreviations used in the column of preparation method represent the following meanings: MSS = microwave solvothermal, CP = cold pressing, SG =

solution grow, SPS = spark plasma sintering, HP = hot pressing, LIE = lithium ionic exfoliation, BMA = ball milling alloy, MA = melting alloy, HPS = high pressure synthesis, BS = Bridgman–Stockbarger, BM = ball milling, THM = travelling heater method, MS = melt spinning, Te-MS = Te rich melt spinning, ZM = zone melting.

## 6 The characteristics of $Bi_2Te_3$ families as thermoelectric applications

### 6.1 Intensive bipolar conduction at relatively high temperature

Bipolar conduction is the excitation of minor charge carriers. In *n*-type semiconductors (as an example), electrons are thermally excited from the valence band to the conduction band, leaving behind holes as the minor charge carriers in the valance, as illustrated by the schematic diagram of Fig. 10.[124] Since electrons and holes have opposite charges, the total $S$ is offset if both electrons and holes are present, refer to Fig. 3b. For semiconductors, $S$ increases with elevating temperature, and the turn-over of $S$ is caused by the bipolar conduction.[124,125] Because the existence of bipolar conduction is related to the band gap ($E_g$), the Goldsmid-Sharp relation (*i.e.* $E_g = 2eS_{max}T$ with $S_{max}$ and $T$ representing the peak value of $S$ and the corresponding temperature, respectively) was proposed to roughly estimate $E_g$ according to the variation of $S$ with temperature.[126] Later on, a more precise method by taking into account the different weighted carrier mobility ratios between balance band and conduction was proposed to estimate $E_g$ based on temperature-dependent $S$.[127] Despite the detrimental effect on $S$, the minor charge carriers also contribute to $\kappa_{bi}$, which has been quantitatively studied previously as shown in Fig. 3f and Fig. 7a. Although $\kappa_{bi}$ is lower than $\kappa_e$ and $\kappa_l$ in most cases, the high-temperature $zT$ is sensitive to $\kappa_{bi}$. Based on above discussion, we can see that the suppression of bipolar conduction can enhance the overall $zT$ from two aspects — shifting the peak of $S$ to high temperature and reducing $\kappa$.

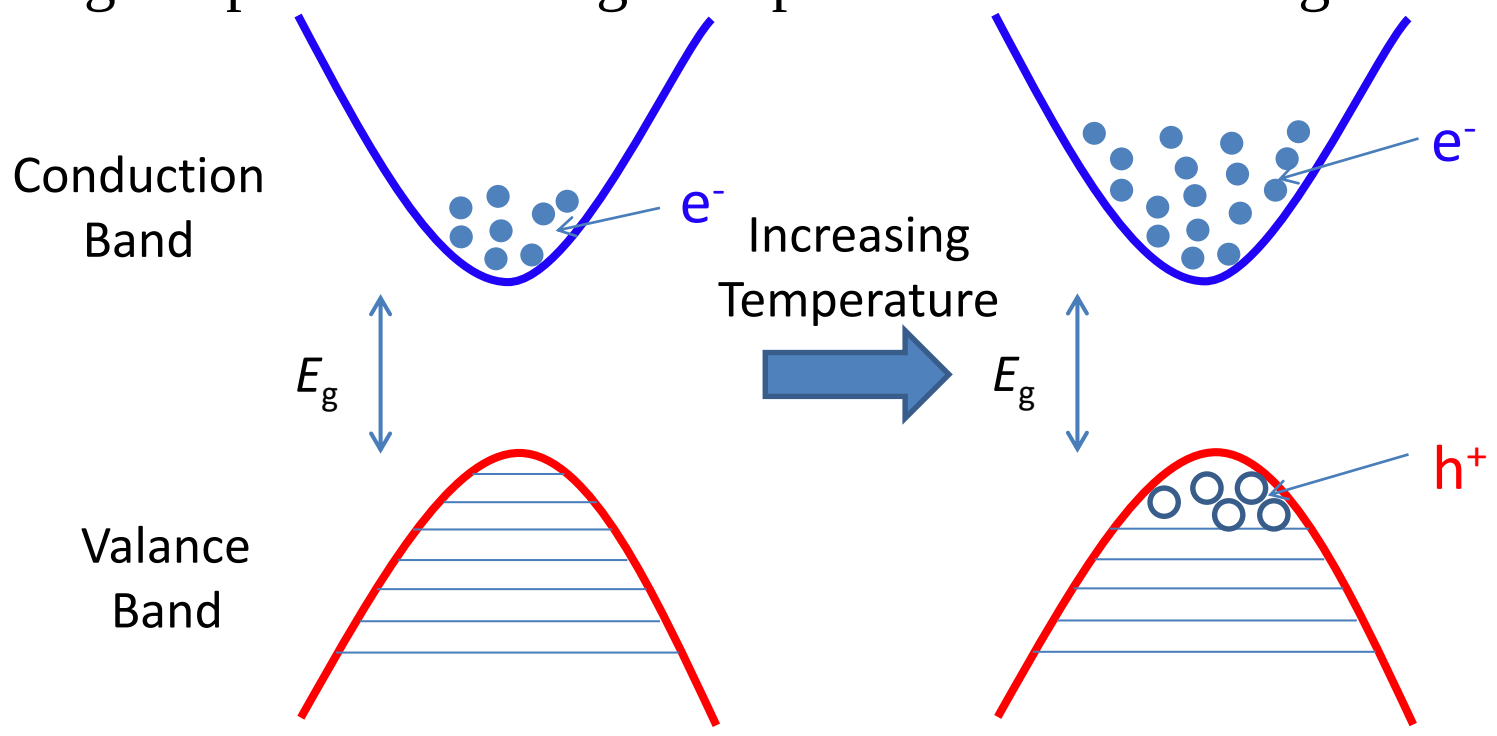

Fig. 10. (color online) Schematic diagram illustrating the bipolar conduction.

As demonstrated in Table 1, $Bi_2Te_3$ families are semiconductors with narrow $E_g$. Consequently, bipolar conduction is prominent at relatively high temperature, which deteriorates their performance. Thereby, it is necessary to suppressing bipolar conduction, which will be discussed in Section 7.

### 6.2 Anti-site defects and vacancies

For $Bi_2Te_3$, the most common defects are vacancies at Te sites and anti-site defects of Bi in Te-sites.[128] The formation of vacancies is caused by the evaporation of consisting elements,[122] and the motivation of anti-site defects is the differences of electronegativity and atomic size between Te and Bi[129]. The formation of Te vacancies (assuming $x$ mol from 1 mol $Bi_2Te_3$) follows

$$Bi_2Te_3 \rightarrow 2Bi_{Bi}^{\times} + (3-x)Te_{Te}^{\times} + xTe(g)\uparrow + xV_{Te}^{2+} + 2xe^{-} \qquad (51)$$

On the basis of $x$ mol $V_{Te}^{2+}$ in 1 mol $Bi_2Te_3$, the generation of $y$ mol anti-site defects of Bi at Te site can be expressed as

$$Bi_2Te_3 \rightarrow (2-z)Bi_{Bi}^{\times} + zBi_{Te}^{-} + (3-x)Te_{Te}^{\times} + xTe(g)\uparrow + (x-z)V_{Te}^{2+} + 2xe^{-} + 3zh^{+} \quad (52)$$

In the above equations, $Bi_{Bi}^{\times}$ and $Te_{Te}^{\times}$ are Bi and Te atoms at their original sites in the lattice, Te(g) is the evaporation of Te, $V_{Te}^{2+}$ is the Te vacancies, and $Bi_{Te}^{-}$ is the anti-site defect of Bi at Te site, respectively. From Equations (51) and (52), we can see that $V_{Te}^{2+}$ is positively charged, and one $V_{Te}^{2+}$ contributes two electrons, whereas $Bi_{Te}^{-}$ is negatively charged, and one $Bi_{Te}^{-}$ gives one hole. Most $Bi_2Te_3$ single crystals or ingots with coarse grains are intrinsically *p*-type because $Bi_{Te}^{-}$ is the dominating defects. For fine-grained polycrystalline samples and nanostructures, the dangling bonds at grain boundaries due to Te deficiencies can also be considered as fractional $V_{Te}^{2+}$, therefore more $V_{Te}^{2+}$ is generated, suggesting fine-grained polycrystalline samples and nanostructures are generally *n*-type.[122]

Likewise, in $Bi_2Se_3$, $Sb_2Se_3$, and $Sb_2Te_3$, there also exist positively charged anion vacancies of $V_{Te}^{2+}$, and $V_{se}^{2+}$, as well as negatively charged anti-site defects of $Bi_{Se}^{-}$, $Sb_{Se}^{-}$, and $Sb_{Te}^{-}$. The formation of anion vacancies depends on the evaporation heat, and the formation of anti-site defects rely on the differences of electronegativity and covalent radius between the consisting cation and anion atoms. Table 3 lists the parameters of evaporation heat, electronegativity and ionic radius of Te, Se, Bi, and Sb. As can be seen, $V_{se}^{2+}$ is more easily to happen than $V_{Te}^{2+}$, and the formation of anti-site defects follows the sequence (easy to difficult) of $Sb_{Te}^{-}$, $Bi_{Te}^{-}$, $Sb_{Se}^{-}$ and $Bi_{Se}^{-}$. That is why single crystal $Sb_2Te_3$ is degenerated *p*-type semiconductor, $Bi_2Te_3$ is nearly intrinsic *p*-type, and $Bi_2Se_3$ is degenerate *n*-type, respectively.

Table 3. The electronegativity and evaporation heat for Te, Se, Bi, and Sb.

| | Te | Se | Bi | Sb |
|---|---|---|---|---|
| Evaporation heat / kJ mol$^{-1}$[130] | 52.55 | 37.70 | 104.8 | 77.14 |
| Electronegativity[129] | 2.1 | 2.55 | 2.02 | 2.05 |
| Covalent radius / Å[130] | 127.6 | 78.96 | 208.98 | 121.75 |

The existence of defects can also enhance the scattering of phonons with high frequency due to the mass fluctuation and strain.[131] However, defects make it hard to tune the thermoelectric properties and lead to the irreproducibility of the obtained high thermoelectric properties.[122] For the nanostructures and ball milling samples, there are more anion vacancies due to the dangling bonds in the dense grain boundaries. Unfortunately, the number of anion vacancies cannot be well controlled.

### 6.3 Strong anisotropic behavior

$Bi_2Te_3$ crystal has remarkable anisotropy that originates from the layered rhombohedral structure. Specifically, the $\sigma$ and $\kappa$ in *a*-*b* plane (perpendicular to the c-axis) are about four and two times, respectively, larger than those along the *c*-axis in $Bi_2Te_3$.[102] The $S$ shows the only slight difference with respect to anisotropy. So, $zT$ in the *a*-*b* plane is approximately two times as large as that along the *c*-axis, as shown in Table 4.[121] On the contrary, thermoelectric properties of $Sb_2Te_3$ single crystal exhibit weaker anisotropic behavior, and the $zT$ values along the two perpendicular directions are nearly identical.

In most cases, we use nanostructuring or ball milling to reduce the grain size for obtaining a significantly reduced $\kappa$. However, this would simultaneously deteriorate $S^2\sigma$ to some extent, resulted from the random crystal orientation. Since $Bi_2Te_3$ shows stronger anisotropic behavior than $Sb_2Te_3$, while the state-of-the-art $zT$ is only ~1.2[100] for $Bi_2Te_3$-based materials, $zT$ in polycrystalline $Sb_2Te_3$-based materials can be up to 1.56[103] because of the dramatically reduced $\kappa$ and the preserved high $S^2\sigma$ (refer to Table 2).

Table 4. The anisotropic behavior of single crystals of $Bi_2Te_3$ families.

| | $\sigma_{\perp c}/\sigma_{\parallel c}$ | $S_{\perp c}/S_{\parallel c}$ | $(S^2\sigma)_{\perp c}/(S^2\sigma)_{\parallel c}$ | $\kappa_{\perp c}/\kappa_{\parallel c}$ | $zT_{\perp c}/zT_{\parallel c}$ |
|---|---|---|---|---|---|
| $Bi_2Te_{2.6}Se_{0.4}$[121] | 4.38 | 1.03 | 4.65 | 2.15 | 2.17 |
| $Bi_{0.5}Sb_{1.5}Te_3$[132] | 2.65 | 1.02 | 2.75 | 1.93 | 1.42 |

## 7 Strategies for enhancing thermoelectric performance of $Bi_2Te_3$ families

Based on the summarized feature of $Bi_2Te_3$ families, some strategies have been developed to enhance the thermoelectric performance. In fact, these strategies are combined together to achieve a higher $zT$.

### 7.1 Suppressing bipolar conduction

As discussed in Section 4, increasing $E_g$ is effective to suppress bipolar conduction. Forming ternary phases of $Bi_2Te_{3-x}Se_x$ and $Bi_xSb_{2-x}Te_3$ can tune $E_g$. Fig. 11a shows the $E_g$ for $Bi_2Te_{3-x}Se_x$ with different compositions. With increasing Se content, $E_g$ for $Bi_2Te_{3-x}Se_x$ increases until x = 1, and then decreases. The reason for the variation of $E_g$ with Se content is due to the chemical bonding environment and the electronegativity difference. As demonstrated earlier, $Bi_2Te_3$ consists of quintuple layers, in which the atoms are arranged in the order of Te(1) – Bi – Te(2) – Bi – Te(1) with two types of differently bonded Te atoms.[133] While chemical bonding between Bi − Te(2) is pure covalent, it is slightly ionic but still covalent in nature between Bi−Te(1), suggesting the bonding between Bi − Te(1) is stronger.[134] To form $Bi_2Te_{3-x}Se_x$, Se atoms firstly substitute Te at Te(2) sites. When x > 1, Se further goes to Te(1) sites randomly. Owing to Se being more electronegative than Te (refer to Table 3),[129] the chemical bonding of Bi – Se is stronger than Bi – Te(2), which enables electrons in $Bi_2Te_{3-x}Se_x$ being more localized. It is understood that with increasing x from 0 to 1 in $Bi_2Te_{3-x}Se_x$, $E_g$ increases. Thereafter, $E_g$ tends to decrease with further increasing x > 1, because Se may weaken Bi – Te(1) bonding.

For $Bi_xSb_{2-x}Te_3$ with an electronegativity of Bi slightly less than that of Sb, it is understood that with increasing Sb content, $E_g$ increases, as depicted in Fig. 11.

Noteworthy, it is well-documented that the optimal compositions corresponding to the peak $zT$ values are $Bi_2Te_{2.7}Se_{0.3}$, and $Bi_{0.5}Sb_{1.5}Te_3$, in which $E_g$ values are not the highest.[135] To increase $zT$, we should simultaneously control the band structures (including $E_g$, $m_b^*$, and $E_f$) and phonon scatterings. Among these strategies, enlarging $E_g$ is still a critical factor in $Bi_2Te_{2.7}Se_{0.3}$, and $Bi_{0.5}Sb_{1.5}Te_3$ to achieve higher $zT$ than their respective binary phases.

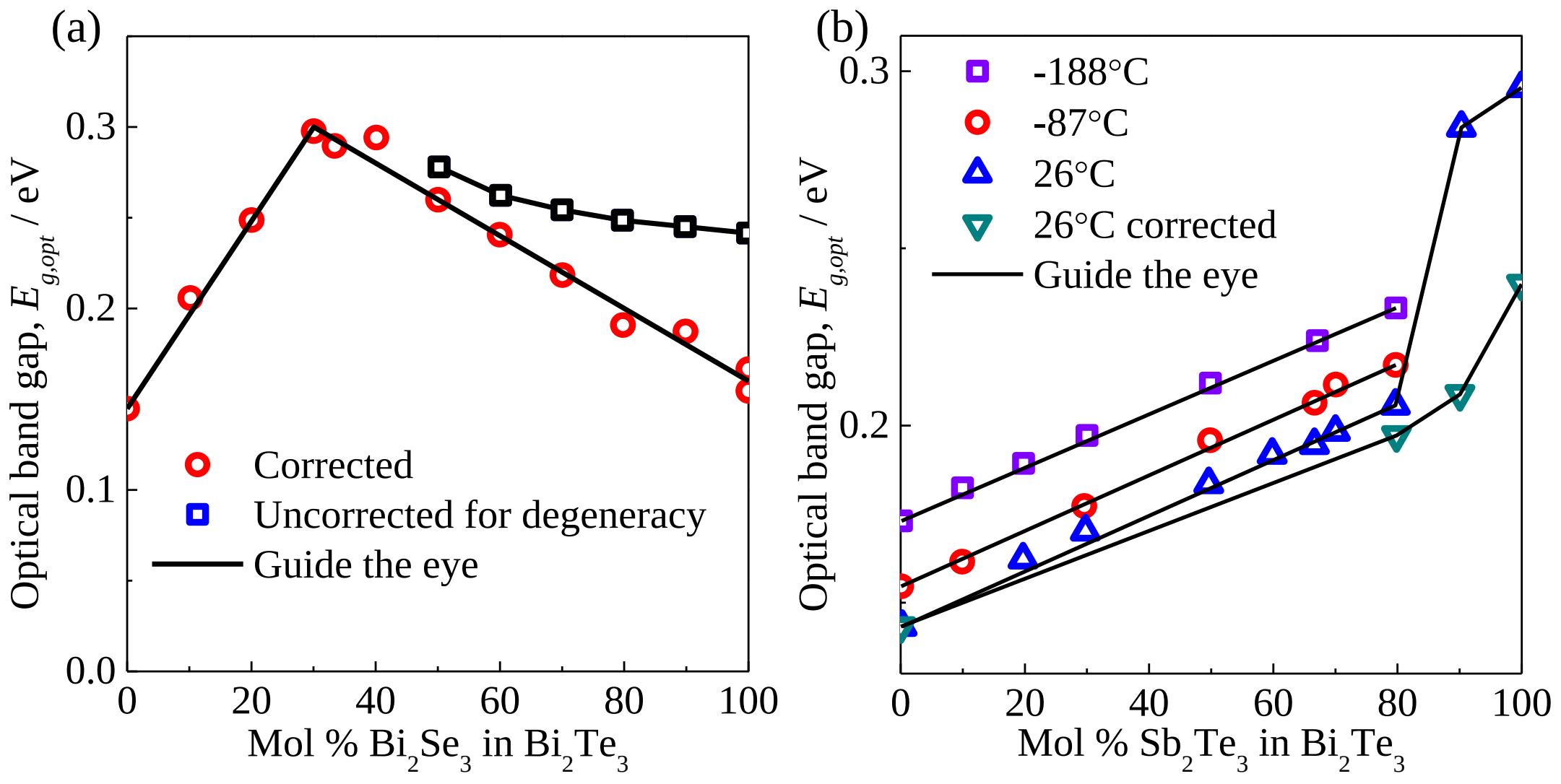

Fig. 11. (color online) Composition-dependent $E_g$ of (a) $Bi_2Te_{3-x}Se_x$ and (b) $Bi_xSb_{2-x}Te_3$. Reproduced from Ref. [136,137].

Moreover, reducing the size of nanostructures can also increase $E_g$ due to the quantum confinement effect.[138-140] For example, it has been confirmed that $E_g$ of $Bi_2Se_3$ nanosheets with a thickness of 1 nm is ~ 0.85 eV, much higher than that of 0.3 eV for the bulk counterpart.[141] Consequently, reducing the thickness of $Bi_2Se_3$ nanosheets can enhance the thermoelectric performance. Experimentally, *zT* of $Bi_2Se_3$ nanosheets with 1 nm was enhanced to be over 0.3.

In respect of suppressing the bipolar conduction, we can also introduce potential barriers to resist the transport of minor carriers. Considering the band alignment between $Bi_2Te_3$ and $Bi_2Se_3$, potential barriers of 0.26 eV in the valence band and 0.11 eV in the conduction band of the *n*-type $Bi_2Te_3/Bi_2Se_3$ could be formed.[142] In this scenario, the transport of excited minor carriers (holes) in the valance band could be resisted.[76] Combined with the energy filtering effect on the major carriers in the conduction band, thermoelectric performance of the *n*-type $Bi_2Te_3/Bi_2Se_3$ was enhanced significantly.[108] Moreover, increasing the carrier concentration to push $E_f$ away from the band gap region can also be effective to suppress bipolar conduction, because the bipolar effect is the strongest in the band gap region (refer to Fig. 3b and f). In nanostructures prepared by solution or ball milling methods, more point defects would be formed, which results in higher carrier concertation.[100] Therefore, we can observe that in these products, the peak *zT* generally occurs at relatively high temperature.[102,112,122,143,144]

### 7.2 Point defect engineering

As mentioned previously, point defects unavoidably present in $Bi_2Te_3$ families, and significantly affect the thermoelectric properties. On one hand, the anion vacancies and antisite defects serve as donors and acceptors, respectively, to determine the carrier type and carrier concentration. On the other hand, point defects, leading to the mass fluctuations and lattice strains, can enhance the scattering of high-frequency phonons to reduce $\kappa$.[131] In this regard, it is necessary to control the inherent vacancies and antisite defects by point defect engineering. The effective methods are mainly to form the ternary phases (*i.e.* n-type $Bi_2Te_{3-x}Se_x$ and p-type $Bi_xSb_{2-x}Te_3$), and doping. In this section, we focus on the effect of point defects on electronic transport. Note that the smaller evaporation energy leads to the easier formation of vacancies, and the smaller differences of electronegativity and atomic size motivate the formation of antisite defects.[145] From Table 3, the point defects strongly depend on the composition of $Bi_2Te_{3-x}Se_x$ and $Bi_xSb_{2-x}Te_3$. Specifically, increasing Se content in $Bi_2Te_{3-x}Se_x$ increases anion vacancies but decreases antisite defects, and increasing Bi content in $Bi_xSb_{2-x}Te_3$ can suppress antisite defects but does not notably affect the anion vacancies.

To clarify the effects of point defects on electronic transport, we summarized the reported data for these ternary phases. Fig. 12 summarized $n_H$, $\mu_H$, $\sigma$, and $S$ at 300 K for the representative *n*-type $Bi_2Te_{3-x}Se_x$ single crystal,[146] single crystal doped with Ag (0.1%),[146] hot pressing plus hot deformation (BM+HP+HD) processed sample,[100] and ingot doped with I (0.08 wt%).[123] From Fig. 12a, with increasing Se content in $Bi_2Te_{3-x}Se_x$, $n_H$ for the single crystals and the ingot gently increases, whereas $n_H$ for the samples prepared by BM+HP+HD initially decreases and then increases. The general increasing trend of $n_H$ with increasing Se content is ascribed to the increase in anion vacancies dominating over that of antisite defects. The decrease in $n_H$ for the BM+HP+HD processed samples with low Se content is caused by the antisite defects exhibiting a stronger effect on carrier concentration over vacancies. From Fig. 12b, $\mu_H$ for single crystals and ingot show intensive fluctuations, while that for BM+HP+HD processed samples stabilize at a plateau, which means the effects of point defects on $\mu_H$ also depends on the sample preparation methods. Because of the variation of $n_H$ and $\mu_H$ caused by point defects, $\sigma$ and $S$ are modified, shown in Fig. 12c and d, respectively. While $\sigma$ generally increases with increasing Se content, $S$ decreases. Based on the discussion in Section 4, large $n_H$ means high $\eta$, resulting in high $\sigma$ but low $S$. Moreover, in the pristine $Bi_2Te_{3-x}Se_x$ single crystal with lower $n_H$ (black data points), its $S$ transfers from *p*-type to *n*-type with increasing Se content, suggesting the major carriers change from holes to electrons. This is because the $Bi_2Te_3$ single crystal is a nearly intrinsic *p*-type semiconductor, whereas substituting Te in the lattice of $Bi_2Te_3$ single crystal with Se atoms can increase anion vacancies contributing more electrons, which produces the transition from *p*-type to *n*-type.

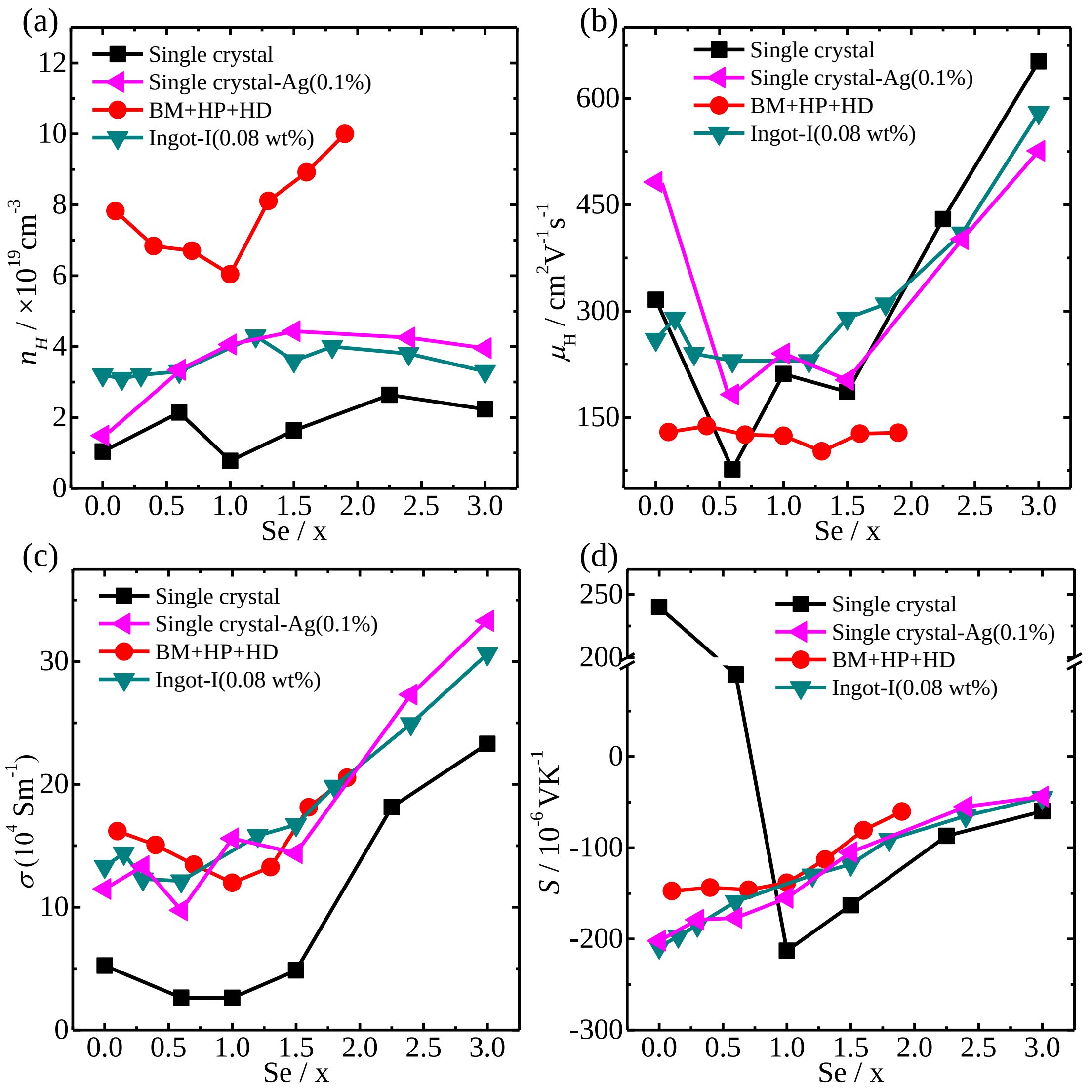

Fig. 12. (color online) $n$-Type $Bi_2Te_{3-x}Se_x$ with (a) $n_H$, (b) $\mu_H$ (c) $\sigma$, and (d) $S$ at 300 K as a function of Se content for single crystal,[146] single crystal doped with Ag (0.1%),[146] BM+HP+HD processed sample,[100] and ingot doped with I (0.08 wt%).[123]

Fig. 13 presents the effect of Bi content on electronic transport properties for the typical $p$-type $Bi_xSb_{2-x}Te_3$ single crystals[132] and BM+HP+HD processed analogs.[100] With increasing Bi content, $n_H$ increases, whereas $\mu_H$ generally decreases. As a consequence, $\sigma$ increases, whereas $S$ generally decreases.

c

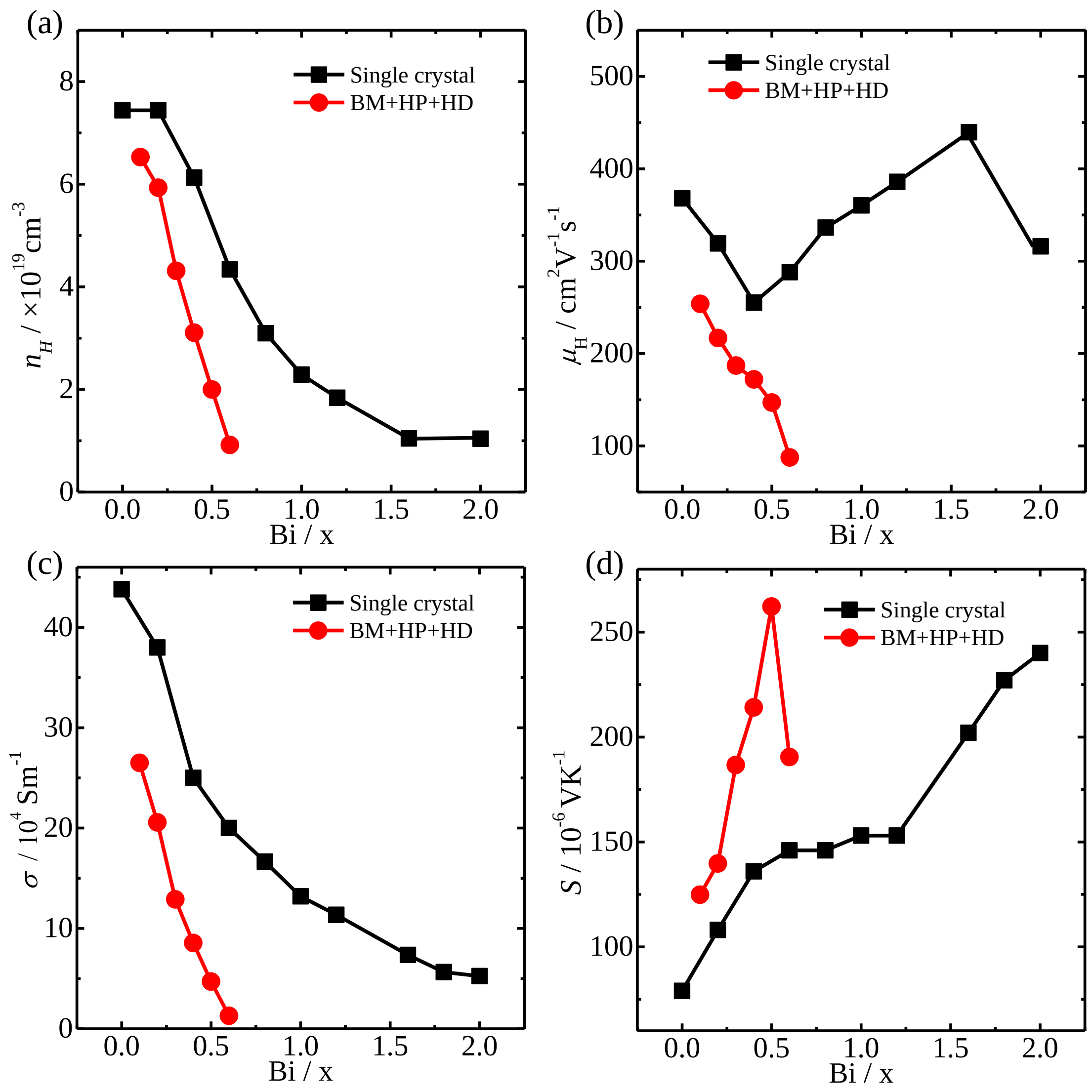

Fig. 13. (color online) *p*-type $Bi_xSb_{2-x}Te_3$ with (a) $n_H$, (b) $\mu_H$ (c) $\sigma$, and (d) $S$ at 300 K as a function of Bi content for single crystal,[132] and BM+HP+HD processed sample.[100]

In addition, doping can also modify the point defects. S-doped $Sb_2Te_3$ can reduce the antisite defects to reduce the carrier concentration and diminish the impurity scattering on holes to enhance the carrier mobility.[115] Another case is Cu-doped $Bi_2Te_{2.7}Se_{0.3}$, which can reduce the vacancies to enhance the carrier mobility.[122] We also noted that hot deformation can reduce the anion vacancies to decrease $n_H$ for the sintered polycrystalline samples. For example, $n_H$ for $Bi_2Te_3$ was reduced to only $1.5\times10^{19}$ cm$^{-3}$ by hot deformation at 733 K, compared with that of $5.9\times10^{19}$ cm$^{-3}$ in the unprocessed counterpart, and as a consequence, $S$ increased from -116 to -141 μV/K.[99,100]

### 7.3 Crystalline alignment

Compared with single crystal $Bi_2Te_3$, polycrystalline materials are promising to realize practical applications, which is due to (1) stronger mechanical strength[147] and (2) dense grain boundaries to reduce $\kappa$.[37] Nevertheless, the strong anisotropic behavior of $Bi_2Te_3$ would worsen $S^2\sigma$ in polycrystalline materials. Therefore, crystalline alignment (*i.e.* enhancing the texture) is likely to enhance $zT$ for polycrystalline samples.[148] Hot deformation is widely used to enhance the texture of $Bi_2Te_3$ families.[149] Fig. 14 summarizes the cutting-edge thermoelectric performance of both *n*-type $Bi_2Te_{3-x}Se_x$ and *p*-type $Bi_xSb_{2-x}Te_3$ benefiting from the hot deformation to enhance the texture of the sintered samples. For *n*-type ones, $\sigma$ is enhanced significantly after hot deformation, while the enhancement of $S$ caused by hot deformation is not so significant. Overall, $S^2\sigma$ for the *n*-type $Bi_2Te_{3-x}Se_x$ is elevated dramatically after hot deformation, which leads to enhanced $zT$. For *p*-type $Bi_xSb_{2-x}Te_3$, hot deformation does not affect $S^2\sigma$ notably but can reduce $\kappa$ to some extent. For this reason, $zT$ of *p*-type candidates can also be enhanced. On this basis, hot deformation enhances $zT$ for *n*-type $Bi_2Te_{3-x}Se_x$ and *p*-type $Bi_xSb_{2-x}Te_3$ from different aspects: enhancing $S^2\sigma$ and reducing $\kappa$, respectively.

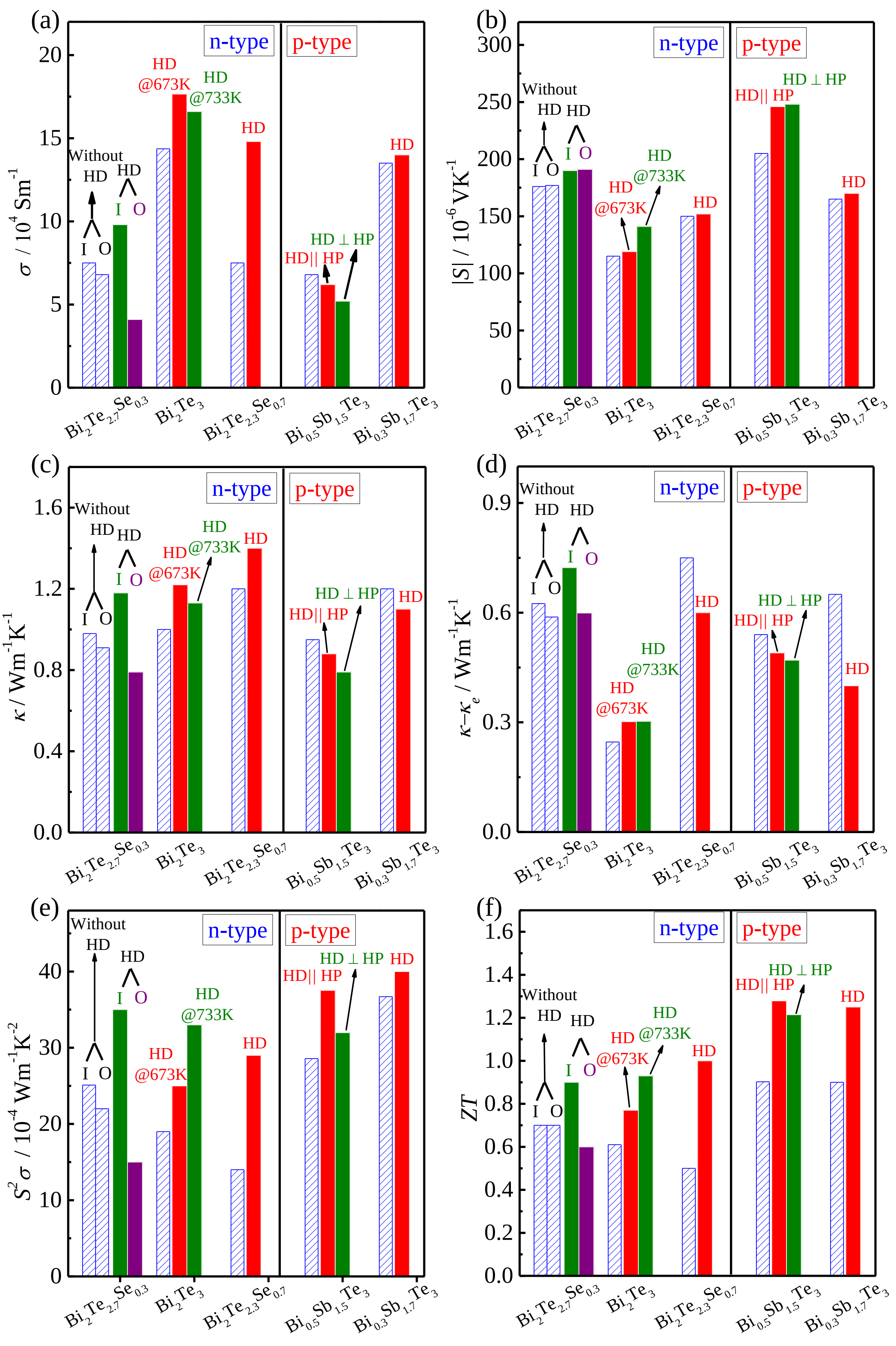

Fig. 14. (color online) The effects of hot deformation on reported (a) $\sigma$, (b) $S$, (c) $\kappa$, (d) $\kappa_l$ (e) $S^2\sigma$, and (f) $ZT$ for n-type $Bi_2Te_{2.7}Se_{0.3}$,[119] $Bi_2Te_3$,[102] $Bi_2Te_{2.3}Se_{0.7}$,[100] and *p*-type $Bi_{0.5}Sb_{1.5}Te_3$,[120] $Bi_{0.3}Sb_{1.7}Te_3$.[100]

### 7.4 Enhancing phonon scattering

Enhancing phonon scattering to reduce $\kappa$ is also effective to enhance the final $zT$. Despite the point defect scattering and hot deformation, which can reduce $\kappa$ (as discussed above), we will cover other strategies for reducing $\kappa$.

#### 7.4.1 Nanostructuring

A variety of one-dimensional, two-dimensional and three-dimensional nanostructures of $Bi_2Te_3$ have been synthesized by solution grow method, including nanowires,[36,150] T-shaped $Bi_2Te_3$-Te heteronanojunctions,[143] Te/$Bi_2Te_3$ nanostring-cluster hierarchical nanostructures,[151-153] hexagonal nanoplates,[106,151,154] $Bi_2Se_3$ ultrathin nanosheets,[113,114,133] $Bi_2Te_3$/$Bi_2Se_3$ multishell nanoplates,[109] and three-dimensional nanoflowers.[108,155] In the synthesis of nanostructures, surfactants are generally used to control the morphology. However, the residuals of surfactants are detrimental to the final thermoelectric performance. Therefore, it is necessary to remove the surfactants or employ the synthesis without any surfactant.

Because of the enhanced phonon scatterings in the nanostructures, $\kappa$ is reduced. Fig. 15a and b show the temperature dependent $\kappa$ and $\kappa - \kappa_e$ for $Bi_{0.5}Sb_{1.5}Te_3$ nanoplates,[38] $Bi_2Te_{2.7}Se_{0.3}$ nanoplates,[38] $Bi_2Se_3$ ultrathin nanosheets,[114] $Bi_2Te_3$/$Bi_2Se_3$ nanoflowers,[108] and $Bi_2Te_3$ nanoplates,[35] compared with the ingot.[123] As can be seen, $\kappa$ of nanostructures can be less than half of the ingot.

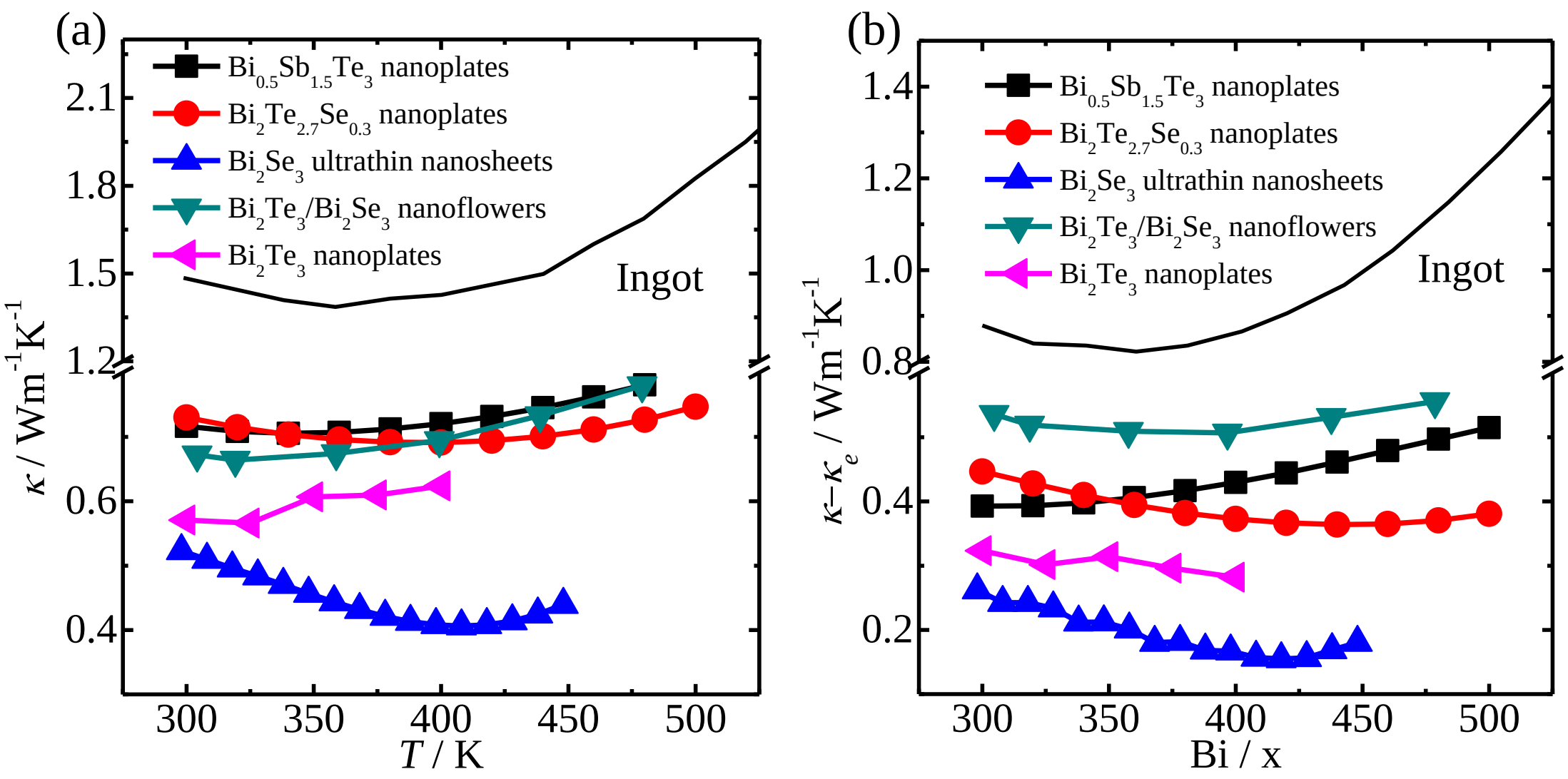

Fig. 15. (color online) (a) $\kappa$, and (b) $\kappa - \kappa_e$ for $Bi_{0.5}Sb_{1.5}Te_3$ nanoplates,[38] $Bi_2Te_{2.7}Se_{0.3}$ nanoplates, $Bi_2Se_3$ ultrathin nanosheets,[114] $Bi_2Te_3$/$Bi_2Se_3$ nanoflowers,[108] and $Bi_2Te_3$ nanoplates,[35] compared with the ingot.[123]

#### 7.4.2 Ball milling

Ball milling can reduce the grain size so as to enhance the grain boundary scattering on phonons. Ball milling generally includes two methods: grounding the ingot with final product phase to obtain fine powders and forming pure phase by high-energy mechanical allying. Both techniques of ball milling have been successfully used to enhance $zT$ for $Bi_2Te_3$ families.[156]

#### 7.4.3 Melt spinning

Melt spinning (MS) can also significantly reduce $\kappa$. Fig. 16 summarizes the thermoelectric performance of $p$-type $Bi_{0.5}Sb_{1.5}Te_3$ prepared by MS.[103,157,158] As can be seen, MS can reduce $\kappa$ by 34% compared with the corresponding ingot and lattice component ($\kappa - \kappa_e$) could be even lower.

In the case of *n*-type ones, MS failed to effectively reduce $\kappa$. Ivanova *et al.* employed MS to prepare *n*-type $Bi_2Te_{2.7}Se_{0.3}$ alloys, but $\kappa$ was not significantly reduced, which resulted in a $zT$ similar to its ingot.[159]

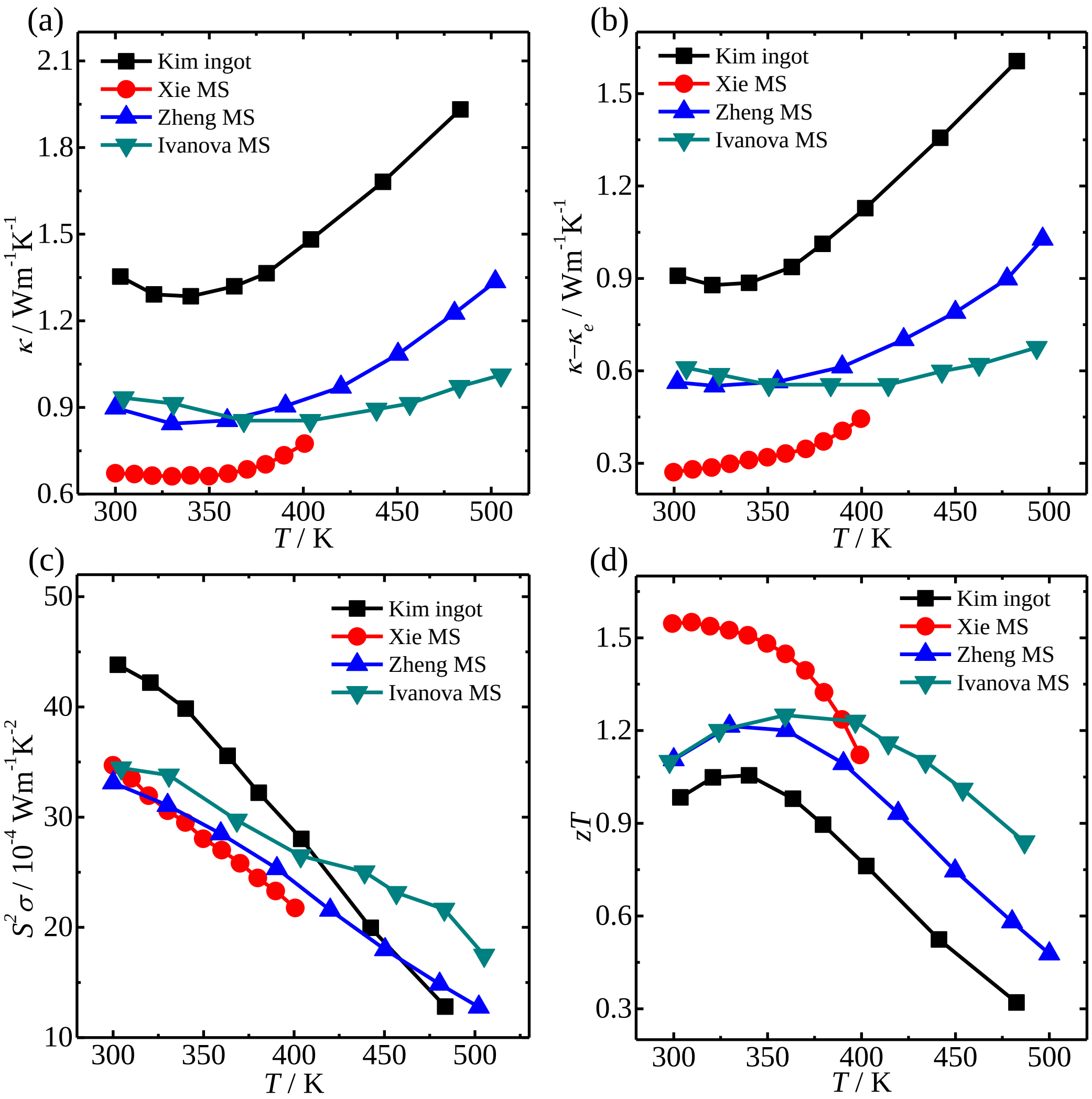

Fig. 16. (color online) (a) $\kappa$, (b) $\kappa_l$, (c) $S^2\sigma$, and (d) $zT$ for $Bi_{0.5}Sb_{1.5}Te_3$ ingot,[160] Xie MS,[103] Zheng MS,[157] and Ivanova MS.[158]

Because thermoelectric properties are correlated with each other, we must employ multi-strategies to enhance thermoelectric performance.[161] Thermoelectric properties are related to the electronic band structure, charge carrier scattering, and phonon scattering. For a specific case with enhanced $zT$ values, it is always due to the combination of multi-strategies to achieve the compromise in these parameters so as to realize a net increase in $zT$.

## 8 Quantitatively Understanding the reported thermoelectric properties

Based on above modeling studies, we will combine the results with the reported thermoelectric properties for $Bi_2Te_3$-based materials to understand the underlying reasons and provide extra hints for further enhancing their performance.

### 8.1 Underlying reasons for the anisotropy behavior of $Bi_2Te_3$ families

Crystallographically, the anisotropy behavior of $Bi_2Te_3$ families is caused by the layered structure. Here, we revealed the underlying fundamentals based on the reported data for single crystals. Fig. 17a and b show the data points of $S$ and $\mu_H$ *versus* $n_H$ for $n$-type $Bi_2Te_{3-x}Se_x$ single crystals, respectively.[121] The curves are theoretical plots of $S$ and $\mu_H$ as a function of $n_H$ calculated with the determined $m_d^*$ for $S$ *versus* $n_H$, and $E_{def}$ for $\mu_H$ *versus* $n_H$. The bold lines correspond to the fitted average values of $m_d^*$ and $E_{def}$. Likewise, Fig. 17c and d present the analysis results for $p$-type $Bi_xSb_{3-x}Te_3$ single crystals.[132] From Fig. 17a and c, the data points of $S$ *versus* $n_H$ for both n-type and $p$-type generally locate near the bold lines, which suggests that the difference of $m_d^*$ along the $ab$-plane and the $c$ axis is small. From Fig. 17b and d, the data points of $\mu_H$ *versus* $n_H$ along the two directions locate near different calculated bold lines. Therefore, the $E_{def}$ values are significantly different along the two directions, and most importantly, such difference in $E_{def}$ values is even larger in $n$-type cases. Fig. 17e and f show the data points of $S^2\sigma$ *versus* $n_H$ and the corresponding theoretical curves for $n$-type and $p$-type ones, respectively. As can be seen, the $S^2\sigma$ along the $ab$-plane is larger than that along the $c$-axis, and such anisotropy is stronger for $n$-type materials.

Based on above discussion, we concluded that the significantly different $E_{def}$ along $ab$-plane and $c$-axis is responsible for the strong anisotropic $\sigma$, and the nearly identical $m_d^*$ leads to the similar $S$ along different directions. In $n$-type case, the difference of $E_{def}$ along the two perpendicular directions is larger than that of $p$-type ones, therefore, anisotropy in $n$-type is even stronger.

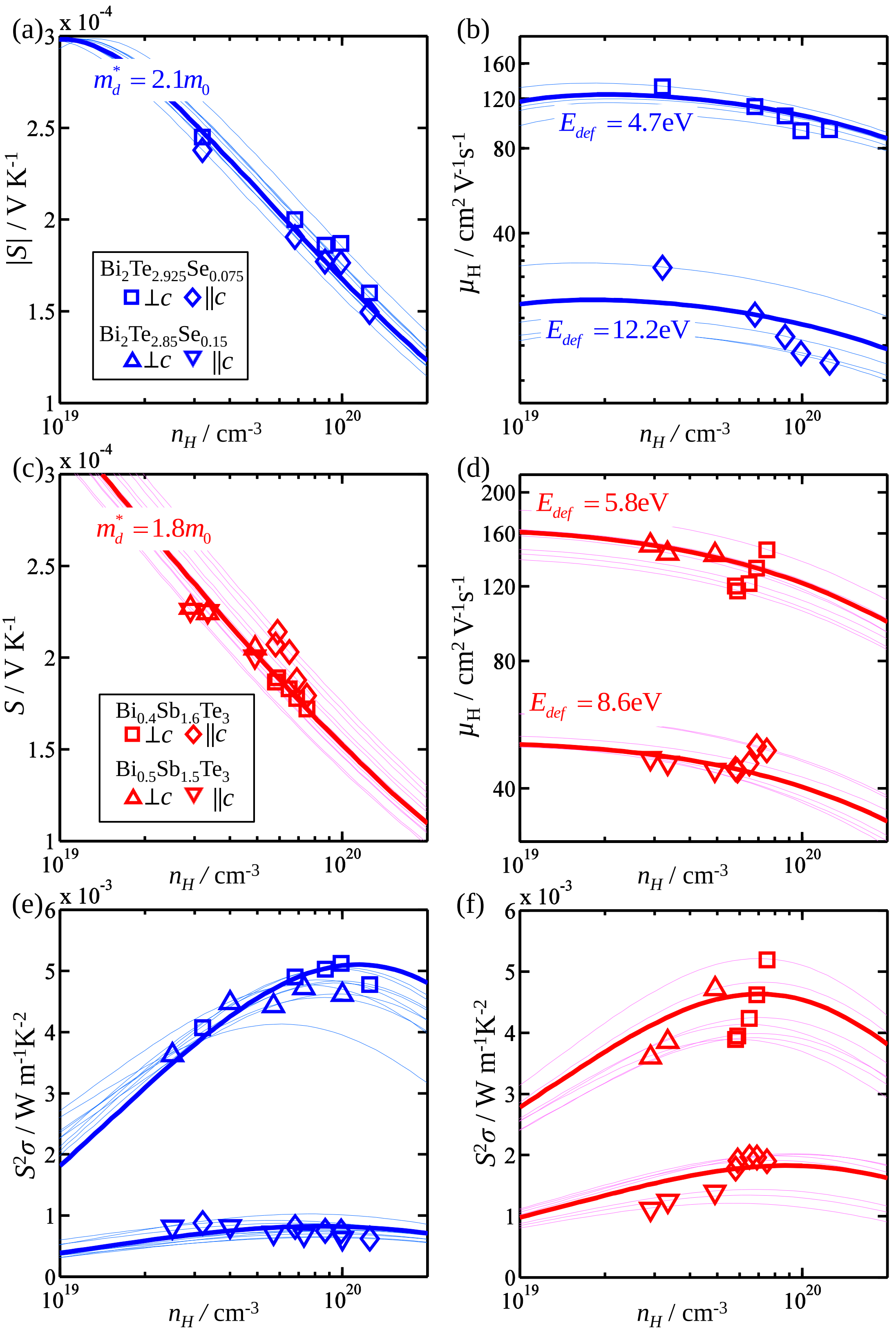

Fig. 17. (color online) (a) Data points of $|S|$ versus $n_H$ compared with the calculated $n_H$ dependent $|S|$ curves, and (b) data points of $\mu_H$ versus $n_H$ compared with the calculated $n_H$ dependent $\mu_H$ curves for $n$-type $Bi_2Te_{3-x}Se_x$ single crystals.[121] (c) Data points of $S$ versus $n_H$ compared with the calculated $n_H$ dependent $S$ curves and (d) data points of $\mu_H$ versus $n_H$ compared with the calculated $n_H$ dependent $\mu_H$ curves for $p$-type $Bi_xSb_{3-x}Te_3$

single crystals.[132] (e) and (f) Data points of $S^2\sigma$ versus $n_H$ compared with the calculated $n_H$ dependent $S^2\sigma$ curves for n-type $Bi_2Te_{3-x}Se_x$ single crystals[121] and *p*-type $Bi_xSb_{3-x}Te_3$ single crystals,[132] respectively.

## 8.2 Understanding the enhanced performance of ternary phases

As discussed above, the enhancement in thermoelectric performance for ternary phases may be caused by the enlarged $E_g$ and point defect engineering. Here, we studied the fundamental reasons for the achieved enhancement in $S^2\sigma$ for the reported *n*-type $Bi_2Te_{3-x}Se_x$. Fig. 18a is the reported Se content dependent data points of $n_H$ for $Bi_2Te_{3-x}Se_x$ ingots doped with I (wt 0.08%) (solid green data points),[123] $Bi_2Te_{3-x}Se_x$ processed by BM+HP+HD (hollow red data points),[100] and $Bi_2Te_{3-x}Se_x$ single crystals (hollow blue data points).[121] Considering the reported $n_H$, $S$, and $\mu_H$ data, we determined $\eta$, $m_d^*$, and $E_{def}$ for all compositions using the Kane band model with CB and VB, shown in Fig. 18b, c, and d, respectively. On this basis, we calculated the theoretical curves of $\mu_H$ and $S$ as a function of $n_H$ for each data point with the correspondingly determined $m_d^*$, and $E_{def}$, exhibited in Fig. 18e and f, respectively, in which the corresponding data points were also presented. The comparison of data points with the theoretical curves suggests that our determinations for $m_d^*$ and $E_{def}$ are sufficiently precise because the data points locate on the relevant curves, and the values of $m_d^*$ and $E_{def}$ strongly depend on the compositions and the material fabrication methods.

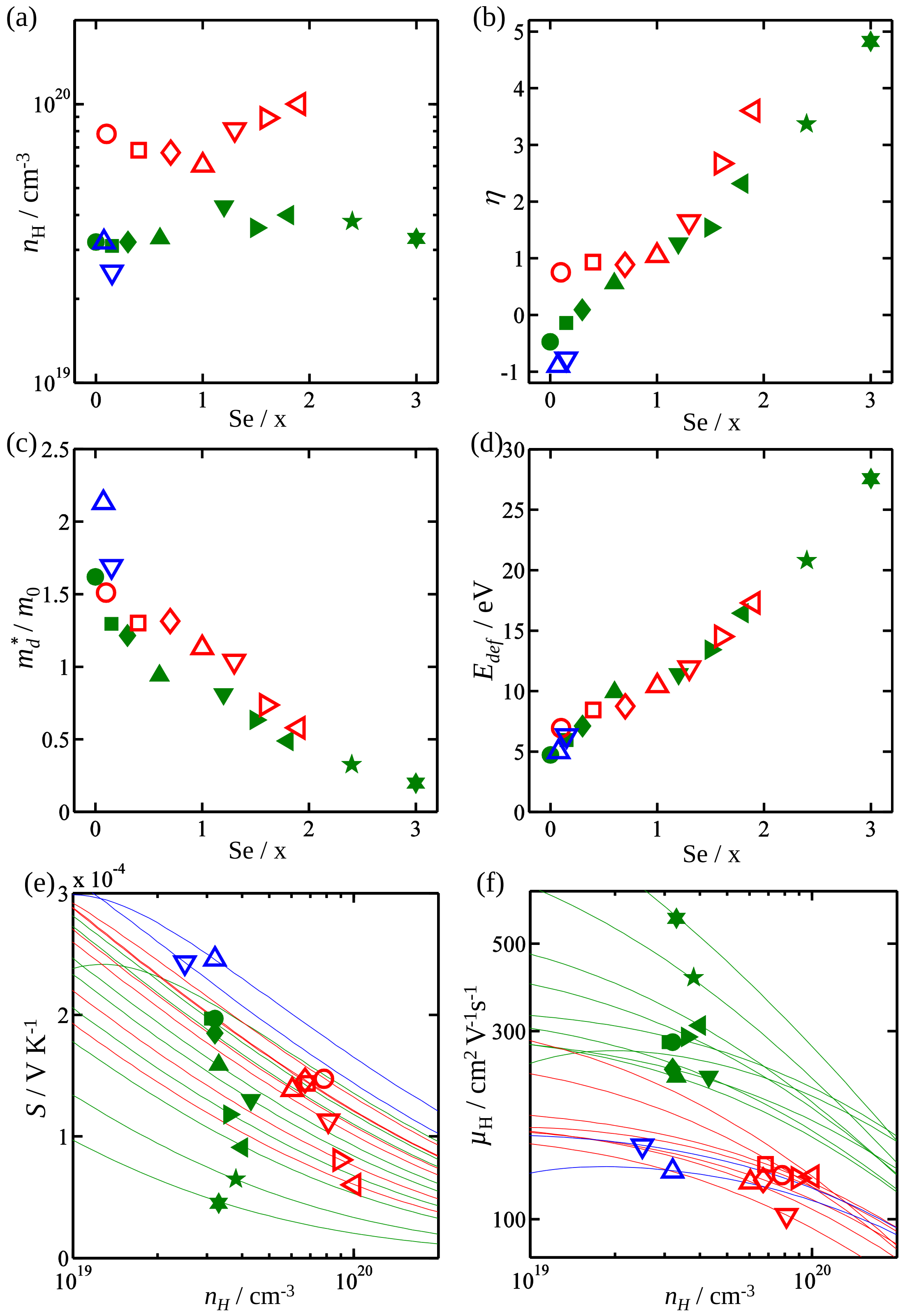

Fig. 18. (color online) The Se content dependent data points of (a) $n_H$, (b) determined $\eta$, (c) determined $m_d^*$, and (d) determined $E_{def}$. The $n_H$ dependent data points of (e) $\mu_H$, and (f) $S$ compared with the theoretical curves of $\mu_H$ versus $n_H$, and $S$ versus $n_H$ calculated with the correspondingly determined $E_{def}$, and $m_d^*$. In all figures, the solid green data points are from $Bi_2Te_{3-x}Se_x$ ingots doped with I (wt 0.08%),[123] the hollow

red data points are from $Bi_2Te_{3-x}Se_x$ processed by BM+HP+HD,[100] and the hollow blue data points are from $Bi_2Te_{3-x}Se_x$ single crystals.[121]

The combination of theoretical curves of $\mu_H$ *versus* $n_H$ and $S$ versus $n_H$ enables to calculate the theoretical curves of $S^2\sigma$ versus $n_H$, shown in Fig. 19a, in which the corresponding data points are also plotted. According to our discussion in Section 4, $n_H^{opt}$ mainly depends on $m_d^*$. To examine the relations between $S^2\sigma$ peak and $n_H$, Fig. 19b demonstrates the data points of determined $m_d^*$ and $n_H$ for all studied materials, compared with the previously determined curve of $m_d^*$ *versus* $n_H^{opt}$ from Fig. 8b. As can be seen, when the data points are located closer to the grey curve in Fig. 19b, it is more likely for the data points in Fig. 19a to approach the peaks of corresponding theoretical curves of $S^2\sigma$ versus $n_H$. However, the difference of $S^2\sigma$ values for different data points cannot be fully explained by the variation of $m_d^*$, although small $m_d^*$ could lead to high $S^2\sigma$. For instance, the green hexagon star with $m_d^*$ of ~$0.2m_0$ shows $S^2\sigma$ of ~$7\times10^{-4}$ $Wm^{-1}K^{-2}$, which is much smaller than $S^2\sigma$ of ~$6\times10^{-3}$ $Wm^{-1}K^{-2}$ for the green round disk with $m_d^*$ of ~$1.7m_0$. Here, we cannot ascribe the enhanced $S^2\sigma$ to the decreased $m_d^*$.

Therefore, we should develop new concepts to understand the enhancement in $S^2\sigma$. As reported in our previous study,[38] we defined the $\lambda E_{def}$ (with $\lambda = (m_I^*/m_0)^{1/2}$), serving as the decoupling factor for $S$ and $\sigma$, and we concluded that reducing $\lambda E_{def}$ is the key to enhance $S^2\sigma$, provided that $\eta$ has been sufficiently optimized. Fig. 19c shows the data points of determined $\lambda E_{def}$ dependent $S^2\sigma$, compared with the theoretical curves of $S^2\sigma$ *versus* $\lambda E_{def}$ in the range of 2 – 6 eV. Note that theoretical curves of $S^2\sigma$ *versus* $\lambda E_{def}$ were calculated according to the optimal $\eta$ ($\eta^{opt}$). Since $\eta^{opt}$ is affected by the $E_g$,[114] for each composition with different $E_g$, there is a unique theoretical curve of $S^2\sigma$ *versus* $\lambda E_{def}$. Considering the small difference of these theoretical curves for various composition, in Fig. 19c we only plotted the representative ones for $Bi_2Te_3$ ingot, $Bi_2Te_{2.4}Se_{0.6}$ ingot, and $Bi_2Se_3$ ingot. Fig. 19d presents the data points of $\lambda E_{def}$ against $\eta$, compared with the $\eta^{opt}$ values, indicated by the vertical lines. As can be seen, small $\lambda E_{def}$ indeed results in large $S^2\sigma$ and narrowing the difference between the determined $\eta$ and the corresponding $\eta^{opt}$ in Fig. 19d allows the data point in Fig. 19c to approach the corresponding theoretical curve.

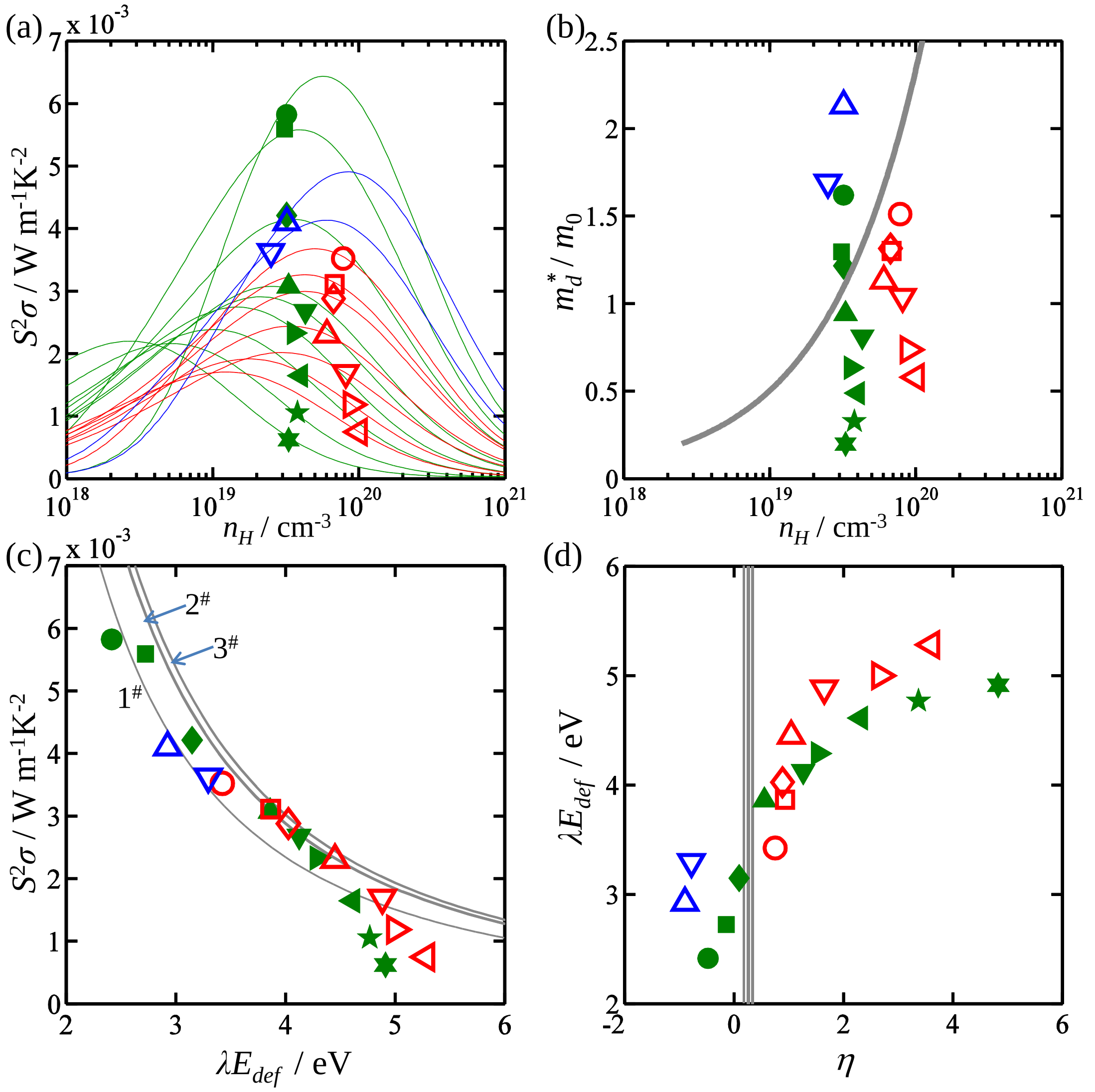

Fig. 19. (color online) (a) Data points of $n_H$ dependent $S^2\sigma$ compared with the theoretical curves of $S^2\sigma$ versus $n_H$ calculated with correspondingly determined $E_{def}$, and $m_d^*$. (b) Determined data points of $n_H$ dependent $m_d^*$ with the grey curve indicating the $m_d^*$ as a function of $n_H^{opt}$. (c) Data points of $\lambda E_{def}$ dependent $S^2\sigma$ compared with the theoretical curves of $S^2\sigma$ versus $\lambda E_{def}$ for compositions of $Bi_2Te_3$ (labeled with 1#), $Bi_2Te_{2.4}Se_{0.6}$ (labeled with 2#) and $Bi_2Se_3$ (labeled with 3#) calculated with the corresponding $\eta^{opt}$. (d) Determined data points of $\lambda E_{def}$ versus $\eta$. In all figures, the solid green data points are from $Bi_2Te_{3-x}Se_x$ ingots doped with I (wt 0.08%),[123] the hollow red data points are from $Bi_2Te_{3-x}Se_x$ processed by BM+HP+HD,[100] and the hollow blue data points are from $Bi_2Te_{3-x}Se_x$ single crystals.[121]

Also, we analyzed the *p*-type BM+HP+HD processed $Bi_xSb_{2-x}Te_3$[100] and the single crystals.[132] Fig. 20a – d show the $n_H$, determined $\eta$, $m_d^*$, and $E_{def}$, respectively. On this basis, we calculated the curves of $S$ and $\mu_H$ as a function of $n_H$ over a wide range using the determined parameters, shown in Fig. 20e and f, respectively.

Fig. 21a and b present the curves of $S^2\sigma$ as a function of $n_H$, and the determined $m_d^*$ *versus* $n_H$. We can observe that when the data point of $m_d^*$ *versus* $n_H$ is close to the optimal curve of $m_d^*$ *versus* $n_H$ in Fig. 21b, the data point of $S^2\sigma$ *versus* $n_H$ is close to the peak of the corresponding curve of $S^2\sigma$ as a function of $n_H$. But, although some data points are close to the peak of the corresponding curve of $S^2\sigma$ as a function of $n_H$, the $S^2\sigma$ could be lower than that for data points deviate from the peak position.

Fig. 21c and d present the $S^2\sigma$ as a function of $\lambda E_{def}$, and $\lambda E_{def}$ *versus* $\eta$, respectively. From which, we can fully understand the reason for enhanced $S^2\sigma$.

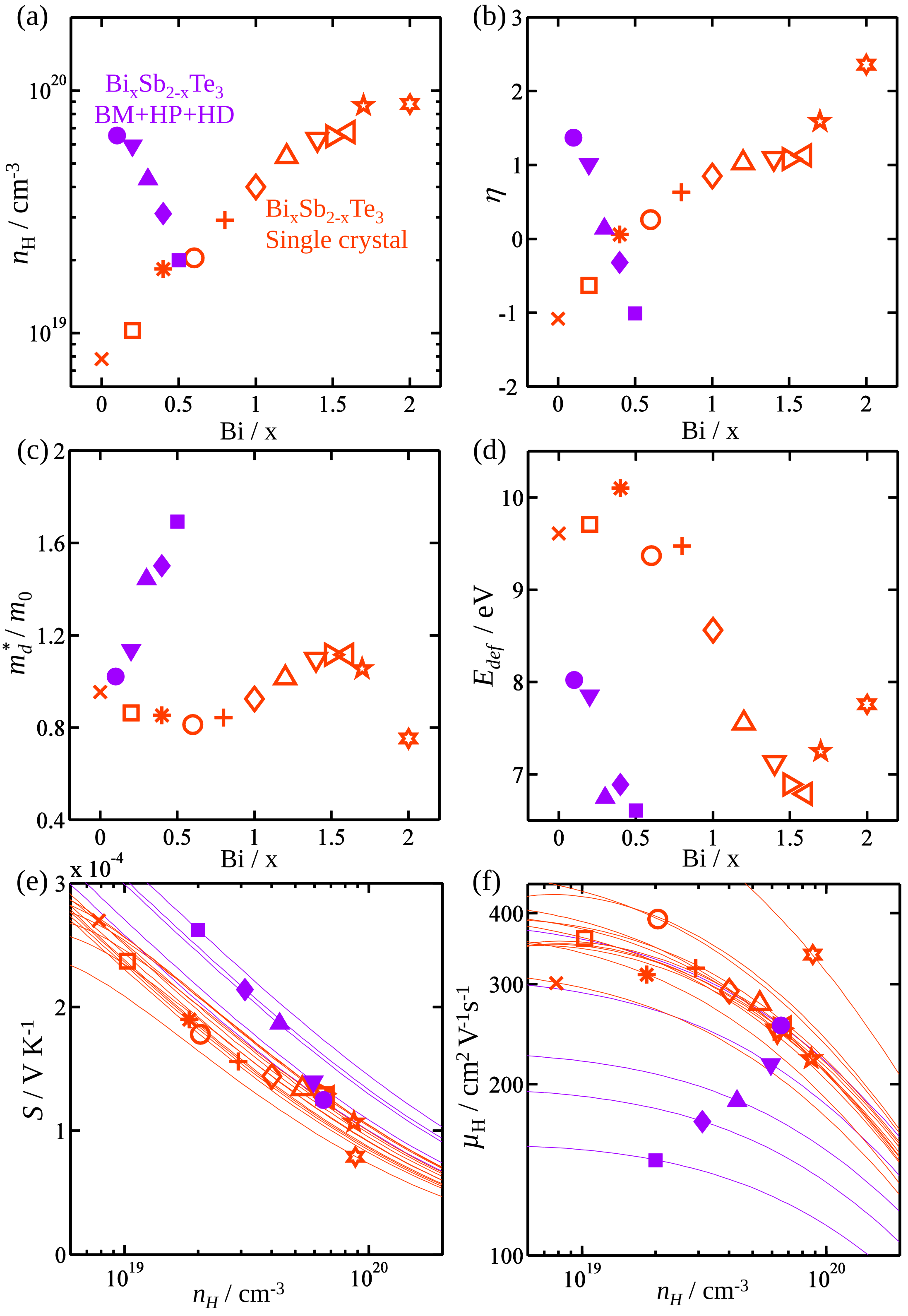

Fig. 20. (color online) (a) $n_H$ of different *p*-type $Bi_xSb_{2-x}Te_3$. Determined (b) $\eta$, (c) $m_d^*$, and (d) $E_{def}$. The $n_H$ dependent data points of (e) $\mu_H$, and (f) $S$ compared with the theoretical curves of $\mu_H$ versus $n_H$, and $S$ versus $n_H$ calculated with the correspondingly determined $E_{def}$, and $m_d^*$. In all figures, the solid purple data points correspond to BM+HP+HD processed samples,[100] and the hollowed data points correspond to single crystals.[132]

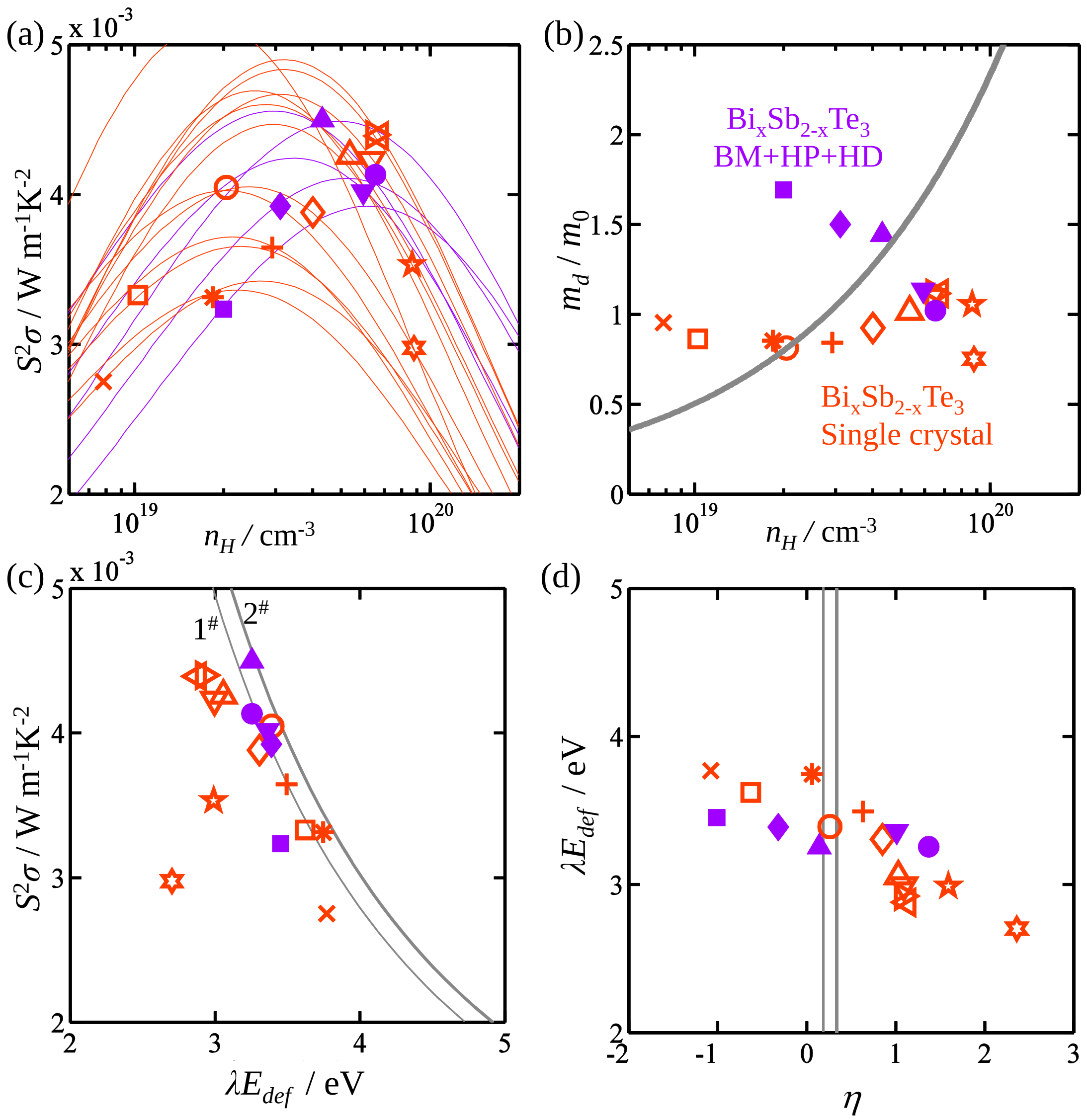

Fig. 21. (color online) (a) Data points of $n_H$ dependent $S^2\sigma$ compared with the theoretical curves of $S^2\sigma$ versus $n_H$ calculated with correspondingly determined $E_{def}$, and $m_d^*$. (b) Determined data points of $n_H$ dependent $m_d^*$ with the grey curve indicating the $m_d^*$ as a function of $n_H^{opt}$. (c) Data points of $\lambda E_{def}$ dependent $S^2\sigma$ compared with the theoretical curves of $S^2\sigma$ versus $\lambda E_{def}$ for compositions of $Bi_2Te_3$ (labeled with 1#), and $Sb_2Te_3$ (labeled with 2#) calculated with the corresponding $\eta^{opt}$. (d) Determined data points of $\lambda E_{def}$ versus $\eta$. In all figures, the solid purple data points correspond to BM+HP+HD processed samples,[100] and the hollowed data points correspond to single crystals.[132]

## 9 Conclusion and perspectives

In this progress report, we studied the physical fundamentals of thermoelectric effects. While Seebeck effect is the distribution of electrons disturbed by temperature difference, and Peltier effect is the energy exchange between free electrons and the surroundings. We also provided a detailed mathematical process of the derivation of electronic transport coefficients from BTE. On this basis, we performed simulation studies on the effects of materials parameters, $m^*$, $E_g$ and $E_{def}$ on these transport coefficients, which provides significant insights into the dependence of thermoelectric properties on materials. We summarized the features of $Bi_2Te_3$ based thermoelectric, and the correspondingly developed strategies for enhancing the thermoelectric performance. Specifically, owing to the narrow band gap, $Bi_2Te_3$ families tend to exhibit intensive bipolar conduction. To suppress bipolar conduction, enlarging band gap is required, which is normally achieved by forming ternary phases or doping. Moreover, minor charge carrier filtering can also be applied to suppress the bipolar conduction. Another feather of $Bi_2Te_3$ is the point defects, which can enhance the mid-frequency phonon scatterings and determine the carrier type as well as carrier concentration. Point defect engineering is widely employed to control the point defects. Finally, caused by the layered structures, $Bi_2Te_3$ shows anisotropic behavior in terms of the thermoelectric properties along the basal plane and the c-axis, which is even stronger in the n-type ones. Therefore, we can observe the *n*-type $S^2\sigma$ is much lower than that of the p-type. Correspondingly, enhancing the texture of polycrystalline materials is anticipated to improve the thermoelectric performance. Moreover, to understand the strong anisotropic behavior and the reported enhanced thermoelectric performance, we re-analyzed the electronic transport properties of $Bi_2Te_3$ based thermoelectric materials. The strong anisotropy is found to be caused by the different $E_{def}$ along the *a*-*b* basal and *c* axis. We further analyzed the achieved enhancement in $Bi_2Te_3$. Based on the modeling studies, we highlight the significance of $\lambda E_{def}$ in enhancing $S^2\sigma$. Reducing $\lambda E_{def}$ is the key to enhance $S^2\sigma$ provided that $\eta$ has been optimized, which is believed to enlighten the development of high-performance thermoelectric materials.

**Acknowledgment**

This work is financially supported by the Australian Research Council. ZGC thanks the USQ start-up grant and strategic research grant.